\documentclass[11pt, a4paper]{article}

\usepackage[margin=1in]{geometry}

\RequirePackage{amsthm,amsmath,amsfonts,amssymb}
\RequirePackage[authoryear]{natbib}
\RequirePackage[colorlinks,citecolor=blue,urlcolor=blue]{hyperref}
\RequirePackage{graphicx}

\usepackage{bm}
\usepackage{bbm}
\usepackage{arydshln}
\usepackage{authblk}
\usepackage{setspace}
\usepackage{enumitem}
\usepackage{rotating}
\allowdisplaybreaks

\usepackage{natbib}
\theoremstyle{plain}

\newtheorem{theorem}{Theorem}[section]

\theoremstyle{remark}

\title{\textbf{Online multiple hypothesis testing}}
\author[a]{David S.\ Robertson}
\author[b]{James M.\ S.\ Wason}
\author[c]{Aaditya Ramdas}

\affil[a]{MRC Biostatistics Unit, University of Cambridge, UK}
\affil[b]{Population Health Sciences Institute, Newcastle University, Newcastle, UK}
\affil[c]{Departments of Statistics and Machine
Learning, Carnegie Mellon University, Pittsburgh,
Pennsylvania, USA}

\date{\vspace{-24pt}}

\begin{document}

\maketitle

\onehalfspacing

\begin{abstract}
Modern data analysis frequently involves large-scale hypothesis testing, which naturally gives rise to the problem of maintaining control of a suitable type~I error rate, such as the false discovery rate (FDR). In many biomedical and technological applications, an additional complexity is that hypotheses are tested in an online manner, one-by-one over time. However, traditional procedures that control the FDR, such as the Benjamini-Hochberg procedure, assume that all $p$-values are available to be tested at a single time point. To address these challenges, a new field of methodology has developed over the past 15~years showing how to control error rates for online multiple hypothesis testing. In this framework, hypotheses arrive in a stream, and at each time point the analyst decides whether to reject the current hypothesis based both on the evidence against it, and on the previous rejection decisions. In this paper, we present a comprehensive exposition of the literature on online error rate control, with a review of key theory as well as a focus on applied examples. We also provide simulation results comparing different online testing algorithms and an up-to-date overview of the many methodological extensions that have been proposed.\\

\end{abstract}


\section{Introduction}
\label{sec:intro}


Large-scale hypothesis testing is now ubiquitous in a variety of biomedical and technological applications. For example, many major technology companies perform tens of thousands of randomised controlled experiments (known as A/B tests) each year to make data-driven decisions about how to improve products~\citep{kohavi2020online}. Meanwhile, in genomics it is now routine to test hundreds of thousands of genetic variants for an association with particular phenotypic trait(s). Even in the setting of randomised controlled trials (RCTs) in medicine, there is a growing push towards the use of ``overarching'' trial frameworks to allow the efficient testing of multiple experimental drugs for multiple patient subpopulations.

Performing a large number of hypothesis tests naturally gives rise to the problem of multiple comparisons~\citep{Tukey1953}: given a collection of multiple hypotheses to be tested, the goal is to distinguish which hypotheses are null and non-null, while controlling a suitable error rate (see Section~\ref{subsec:error_rates}). This error rate is generally formed around the probability of incorrectly classifying a null hypothesis as non-null. Typically, a $p$-value is calculated for each hypothesis and is then used to decide whether to reject the null hypothesis. Multiple hypothesis testing is one of the core problems in statistical inference, and has led to a wide range of procedures that can be used to correct for multiplicity and ensure that a suitable error rate is controlled. In contrast, uncorrected hypothesis testing contributes to serious concerns over reproducibility, publication bias and `p-hacking' in scientific research~\citep{Ioannidis2005, Head2015}. 

{Multiplicity, as broadly understood, is naturally linked to scientific reproducibility. \citet{goodman2016does} state that ``Multiplicity, combined with incomplete reporting, might be the single largest contributor to the phenomenon of nonreproducibility, or falsity, of published claims'' and go on to say that ``Scientific fields that routinely work with multiple hypotheses without correcting for or reporting the occurrence of multiplicity run a higher risk of nonreproducibility of results or inferences". As an example of this, \citet{zeevi2020ignored} recently showed that adjusting for multiplicity greatly enhances the reproducibility of results from psychology experiments.  Similarly, in the context of drug development, \citet{bretz2014multiplicity} in a paper titled ``Multiplicity and replicability: two sides of the same coin'' showed that there is a close link between between multiplicity and replicability in terms of the observed effect sizes of selected subgroups, with further examples given in~\citet{bretz2019replicability}.}

Traditionally, multiple hypothesis testing is \textit{offline} in nature, in the sense that a procedure for testing~$N$ hypotheses will receive all of the corresponding $p$-values $(P_1, \ldots, P_N)$ at once. Step-up and step-down multiple testing procedures (for example) require knowledge about all $p$-values in advance. In the offline setting, the seminal Benjamini-Hochberg (BH) procedure is the dominant method used for FDR control. However, this paradigm is often incompatible with modern data-driven decision-making processes, as demonstrated by our motivating examples in Section~\ref{subsec:motivation}. Once the data is available to make a decision about a particular hypothesis, it can be desirable to take a corresponding action (e.g.\ to update a tech product) relatively quickly, and not to wait for the results of ongoing or future hypothesis tests. Linked with this, in many application areas it may not even be possible to know in advance how many tests in total will be performed. Moreover, the repeated application of traditional offline multiple testing procedures as the family of hypotheses grows can lead to repeatedly changing past decisions, which may be undesirable in some contexts.

What is needed therefore are procedures for \textit{online} multiple hypothesis testing, which better take into account the nature of modern data analysis. This is defined as follows:
{A stream of hypotheses arrives online. At each step, the analyst must decide whether to reject the current null hypothesis without having access to the number of hypotheses (potentially infinite) or any future data, but solely based on the previous decisions and evidence against the current hypothesis.} \\


\noindent \textit{{Online and sequential testing}} \\
Online hypothesis testing has a sequential nature, in the sense that individual hypotheses (or batches of hypotheses) are tested one after the other over time. However, this is distinct from the more traditional concept of \textit{sequential testing}, which refers to the testing of a \textit{single} hypothesis in a sequential manner {with data accumulating over time}. In sequential testing, the sample size for the experiment is not fixed in advance, and the accumulating data is evaluated as they are collected to allow the experiment to be stopped adaptively, such as when statistical significance is achieved. The framework of online multiple testing can be expanded to be ``doubly sequential'', where the inner sequential process is a single sequential test, and the outer sequential process refers to the multiple experiments that are performed to test different hypotheses. 

For each null hypothesis $H_t$, an anytime-valid $p$-value is a sequence of $p$-values $(P_{t,n})_{n \geq 1}$ where $n$ indexes the sample size in the experiment corresponding to hypothesis $H_t$, such that $\text{Pr}(P_{t,N} \leq x) \leq x$, for all $x \in [0,1]$ and \textit{any} data-dependent stopping time $N$. In other words, the stopped anytime $p$-value is a valid $p$-value in the classical sense, no matter how the experiment was stopped. In online multiple testing, we typically drop the second index and focus on the ``outer sequential process'' (across experiments/hypotheses), which means that we assume that for each hypothesis $H_t$, we have a valid $p$-value $P_t$, but we keep in mind that this could have been achieved by stopping an anytime-valid $p$-value (the ``inner sequential process'', corresponding to the evidence within a single experiment).

We also wish to draw a distinction between online hypothesis testing and \textit{multi-armed bandit (MAB)} testing. While both frameworks allow the comparison of multiple experimental arms over time, an MAB can be considered as a \textit{single} experiment in which resources are iteratively allocated to the different arms in order to adaptively trade off certain costs and benefits, and this allocation depends on the previously observed outcomes on each arm. Again, the two testing frameworks can be combined within a doubly sequential framework where there is a sequence of MAB problems over time, see~\cite{Yang2017}.

\begin{figure}[ht!]
\centering
\includegraphics[width=0.7\linewidth]{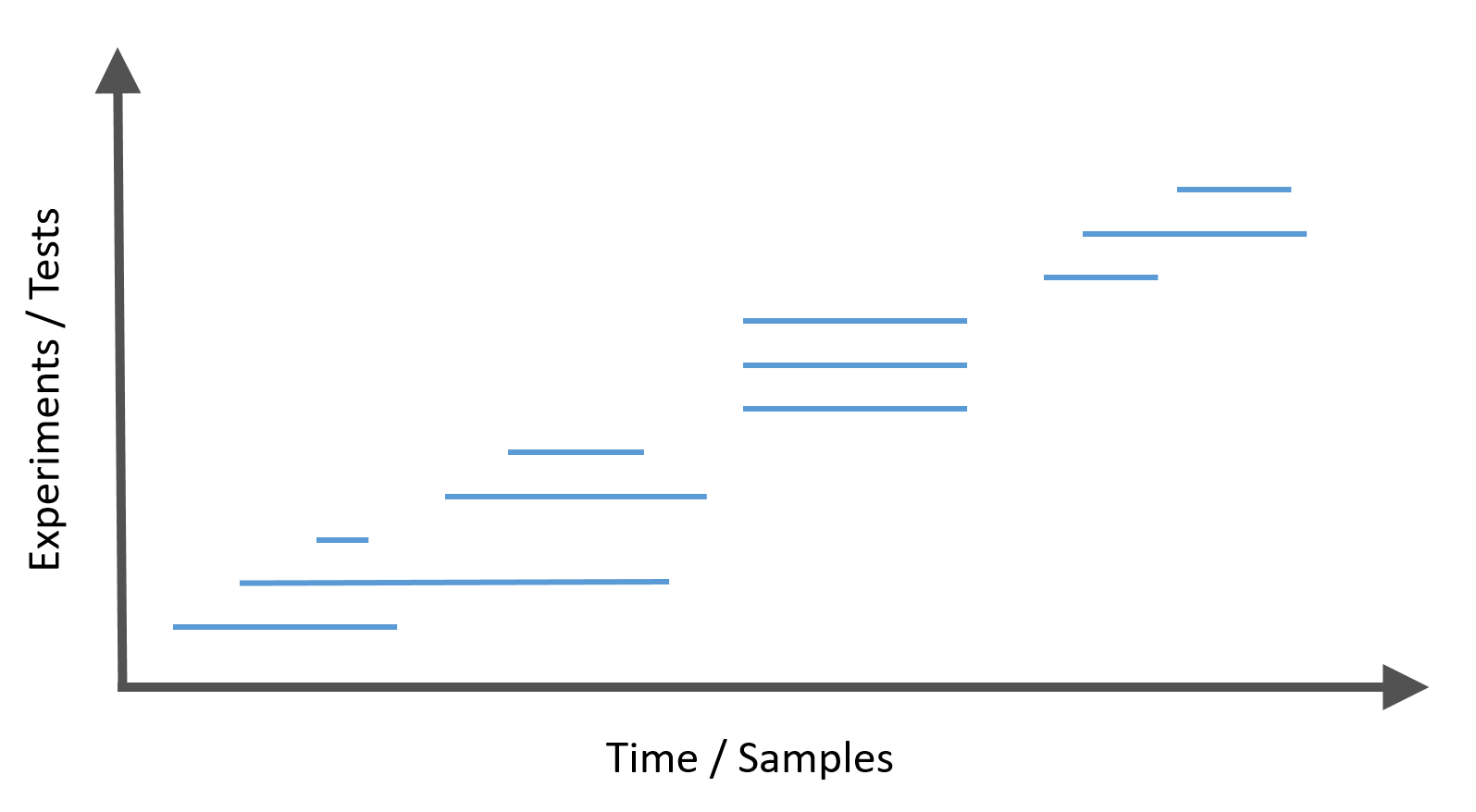}
\caption{An abstract online multiple testing framework. As time passes (left to right), new experiments testing different hypotheses are started and stopped, in a possibly indefinite manner. Each horizontal line represents a new experiment/hypothesis, and the length represents the number of samples collected. Decisions about each hypothesis must be made as soon as the corresponding experiment ends.}
\label{fig:online_testing}
\end{figure}

Figure~\ref{fig:online_testing} gives a diagrammatic representation of online multiple testing, where different hypotheses (corresponding to experiments) are tested over time (corresponding to the collection of data samples). As discussed above, each experiment could itself be a sequential experiment or take the form of an MAB. 

Since the framework was first proposed by~\cite{Foster2008}, a variety of procedures that control error rates for online hypothesis testing have been developed~\citep{Aharoni2014, Javanmard2015,Ramdas2018}. Our aim in this paper is to provide an expository overview of this literature on online error rate control, with a review of the underlying theory, key methods and applied examples.

{The bulk of the literature has focused on the setting of independent hypothesis tests for provable FDR control (with slightly weaker conditions allowing for control of variants of the FDR). Another important feature of many of the algorithms presented here is that they are \textit{adaptive}: when some fraction of the tests actually have the alternative hypothesis true (as evidenced by $p$-values), they adapt and use less conservative tests.}

In the rest of this section, we give motivating examples for online multiple testing and present formal definitions of error rate control. Section~\ref{sec:methods} describes the key procedures for online error rate control in detail. Section~\ref{sec:simulation} presents a simulation study of the procedures, while Section~\ref{sec:case_studies} presents two case studies of applying online error rate control. We describe further methodological extensions as well as future directions in Section~\ref{sec:extensions}. {In Section~\ref{sec:summary} we provide a summary and some  practical guidance,} and conclude with a discussion in Section~\ref{sec:discuss}.

\subsection{Motivating examples}
\label{subsec:motivation}

{We now present three motivating examples from a spectrum of `easier' to `harder' settings for currently available online testing algorithms, in terms of the statistical dependency between hypotheses.} \\


\noindent\textit{A/B testing in tech companies {(independent hypotheses)}}

The development of web applications and services in the tech industry increasingly relies on the use of randomised controlled experiments known as A/B tests. There are a number of widely used platforms now available that streamline and handle the implementation of A/B tests. A typical application is in the development of different versions of webpages. As described in~\citet{berman2021false}, in this context there are two webpage variations (A and B). When an online user visits, the platform randomly assigns the visitor to one of the variations for the duration of the experiment. The platform records the actions that the visitor takes, where the monitored actions reflect the experimenter's goal(s), such as increasing visitor engagement (defined appropriately) or increasing revenue. One of the variations is designated as the baseline, and the performance of the other variation is compared to the baseline using suitable test statistics. If run correctly using anytime-valid $p$-values \citep{johari2021always} and/or confidence sequences~\citep{howard2021time}, the data can be continuously monitored by the experimenter, and a decision can be made at any stopping time.

Many tech companies run tens of thousands of A/B tests per year, as part of a continuous process of designing, delivering, monitoring and improving webpages and other web services. However, there is reason to reduce the number of false alarms which result in making changes to web products that do no better (or even perform worse) than the current iteration, corresponding to an incorrect rejection of the null, particularly when such changes are potentially costly or disruptive to users. Hence, the framework of online error rate control provides a framework to do so while still allowing a large number of A/B tests to be performed in a flexible manner. \\

\noindent \textit{Platform trials {(known positive dependence)}}

A platform trial has a single master protocol that evaluates multiple treatments across one or more patient types, and allows a potentially large number of treatments to be added during the course of the trial~\citep{Saville2016}. {A new treatment can be added to the trial (corresponding to testing a new hypothesis) when a new experimental therapy becomes available, such as when a safe drug candidate for the disease in question is identified from a successful phase~I clinical trial}. Treatments are dropped from the trial after they have been formally tested for effectiveness. Such a trial could (in theory) {be} `perpetual' in that new treatments can continue to enter into the trial and be tested. Figure~\ref{fig:STAMPEDE} in Section~\ref{subsec:STAMPEDE} gives a diagram of an example platform trial showing what this looks like.

In a platform trial, treatments are introduced at different time points by design. However, the trial investigators will wish to make a decision on whether a treatment is beneficial as soon as the data are ready, without waiting for results from the other treatment arms.
%
Hence, the treatment effects are tested sequentially in an online manner, where the number of treatments to be tested in the future may be unknown. More formally, a platform trial generates a sequence of null hypotheses $(H_1, H_2, H_3, \ldots)$ which are tested sequentially. Hypothesis $H_i$ tests the value of some parameter~$\theta_i$, such as an estimate of the treatment difference compared to a control arm.

%
The $p$-values generated from the platform trial described above will not be independent in general. Dependencies will primarily arise due to the shared control data that is re-used to test multiple hypotheses. 
%
%
A current example of a long-running platform trial is the STAMPEDE trial~\citep{james2008stampede} for patients with locally advanced or metastatic prostate cancer, which we return to as a case study in Section~\ref{subsec:STAMPEDE}. \\

\noindent \textit{Data repositories {(unknown arbitrary dependence)}}

Public databases and shared data resources are becoming increasingly pervasive and important in modern biomedical research, particularly in the fields of genetics, molecular biology and routinely collected healthcare records. Some well-known examples include the 1000 Genomes Project~\citep{1000genomes2015}
and the Wellcome Trust Case Control Consortium~\citep{WTCC2007}.
Another example is the International Mouse Phenotyping Consortium database~\citep{Koscielny2013, Dickinson2016}, which we describe as one of our case studies in Section~\ref{sec:case_studies}. Meanwhile, the increase in routinely collected healthcare data allows evaluation of different healthcare technologies  used in practice through emulation of target trials \citep{dickerman2019avoidable}.

Multiple testing naturally occurs in this setting in two ways. Firstly, such databases can be accessed by multiple independent researchers at different times. When a researcher or research group comes up with a new hypothesis, they can fetch the relevant data from a database and perform a statistical test.
Secondly, in {some} databases the family of hypotheses to be tested grows over time as new {new hypotheses are tested (e.g., corresponding to new experiments being performed that measure phenotype expression for a previously untested gene knockout}. In both of these scenarios, the number of hypotheses being tested will be unknown and potentially very large, and lead to concern about overlapping data allowing for arbitrary correlation patterns between hypothesis tests. The issues such dependence causes will be considered throughout the rest of this paper. 

In order to control the number or proportion of false discoveries in this context, new procedures are required that allow a researcher to decide whether to reject a current hypothesis with minimal information about previous hypotheses, and without prior knowledge of even the number of hypotheses that are going to be tested in the future. This is precisely the online multiple testing framework described earlier.

\subsection{Error rates}
\label{subsec:error_rates}


We now formally define some error rates of interest. The basic problem setup is as follows. At each time step $t = 1, 2, \ldots$ the experimenter observes a $p$-value $P_t$ corresponding to {testing} a null hypothesis $H_t$, and must make a decision whether to reject $H_t$ before the next time step. We assume that all $p$-values are valid, i.e.\ if the null hypothesis $H_t$ is true, then $\text{Pr}\{ P_t \leq x\} \leq x$ for all $x \in [0,1]$.\footnote{{Note that if violations of this condition are not too large in the sense that there exists some $\epsilon > 0$ such that
\[
\Pr(P_t \leq x) \leq x(1 +\epsilon),
\]
for every $x \in [0,1]$ then it is straightforward to check that the FDR proofs still go through for the online algorithms presented in Section~2.2, guaranteeing an FDR of $\alpha(1 +\epsilon)$ instead.}} At time $t=0$, the experimenter fixes the level~$\alpha$ at which a suitable error rate is meant to be controlled at all times.

A general testing procedure provides a sequence of test levels~$\alpha_t$ with decision rule 
\begin{equation}\label{eq:general-rejection}
R_t = \begin{cases}
1 \qquad \text{if } P_t \leq \alpha_t \qquad \text{(reject } H_t) \\
0 \qquad \text{otherwise} \qquad \text{(accept } H_t)
\end{cases} ~ .
\end{equation}
At any time~$T$, let $R(T) = \sum_{t=1}^T R_t$ denote the number of rejections (also known as discoveries) made so far and $V(T)$ denote the total number of falsely rejected hypotheses (also known as false discoveries).

The \textit{false discovery proportion} {(FDP)} up to time~$T$ is defined as \[ \text{FDP}(T) := \frac{V(T)}{R(T) \vee 1}, \]
where $a \vee b = \max(a,b)$.
The \textit{false discovery rate} {(FDR)} is then the expectation of the FDP: \[ \text{FDR}(T) := \mathbb{E}\! \left[\frac{V(T)}{R(T) \vee 1}\right].\]
A commonly studied variant is the \textit{marginal} FDR {(mFDR)}: 
\[ \text{mFDR}(T) := \frac{\mathbb{E}[V(T)]}{\mathbb{E}[R(T) \vee 1]}.\]
Another related error rate is the \textit{false discovery exceedance} (FDX), which is the probability the supremum of the FDP exceeds a predefined
threshold~$\epsilon$: \[ \text{FDX}_{\epsilon}(T) := \text{Pr} \left[\text{sup}_{0 \leq t \leq T}\, \text{FDP}(t) \geq \epsilon \right]. \]

{We view the FDR as the central metric of interest, given its long history, widespread use in applied fields such as genetics, and intuitive interpretation. The mFDR can be a convenient theoretically tractable proxy for the FDR when it is not possible to prove FDR control for a particular algorithm and data application, as we highlight in the rest of the paper. In some settings previously explored in the literature, it has been shown empirically that the realised FDR and mFDR of online hypothesis testing algorithms are very similar (see e.g.\ Appendix~F of~\citet{Zrnic2018}), although this is not true in general (see e.g.\ the Supplementary material in~\citet{Javanmard2018}). Hence, the FDR would be the default choice for most users, with the mFDR then being the pragmatic alternative error rate choice if a suitable algorithm for FDR control is not available for the particular data application in mind.}

{In contrast, the FDX gives a stricter guarantee about the distribution of the FDP: whereas the FDR controls the expectation of the FDP, the FDX controls the tail probability of the FDP (i.e., controlling the $(1-\epsilon)$-quantile of the FDP distribution). Control of the FDX makes most sense in settings where the FDP can deviate significantly from its expectation, such as when the number of hypotheses to be tested is not very large, or there is significant correlation~\citep{Javanmard2018}. As for the choice of~$\epsilon$ for the FDX, a default choice of $\epsilon = 0.05$ or $0.10$ is one option, but a pre-hoc choice of $\epsilon$ may also be motivated on practical grounds, such as choosing~$\epsilon$ based on the required sample size to achieve a desired power given control of the FDX at level~$\alpha$. Another approach is to use recently proven post-hoc bounds of the $\text{FDX}_{\epsilon}$ for online testing algorithms under independence, which allows the user to choose~$\epsilon$ (and $\alpha$) freely by examining the corresponding rejections and seeing what makes most sense \citep{katsevich2020simultaneous}}.

An alternative error rate to those based on the FDP is the \textit{familywise error rate} (FWER), which is more commonly considered in clinical trial contexts {due to the relatively small number of hypotheses and regulatory requirements}. The FWER is the probability of falsely rejecting any null hypothesis:
\[
\text{FWER}(T) := \text{Pr} \left[ V(T) \geq 1 \right]. 
\]
The FWER and hence the FDR can be controlled at level~$\alpha$ in a simple manner by using a Bonferonni-type correction, also known as \textit{alpha-spending}. More precisely, we can choose significance levels~$\alpha_t$ for~$H_t$, such that $\sum_{t=1}^{\infty} \alpha_t = \alpha$. {We reiterate that this corresponds to the setting where each nominal critical value $\alpha_t$ corresponds to testing a single hypothesis~$H_t$, with the possibility of repeated testing of the same hypothesis (as in the sequential testing literature) included implicitly, see our remark in the Introduction}.
However, {alpha-spending} suffers from a low statistical power, with the probability of the null hypothesis~$H_t$ being rejected tending to zero as $t$ increases. This motivates the development of more sophisticated algorithms for online error rate control.


\section{Online error rate control methodology}
\label{sec:methods}

\subsection{Generalised alpha-investing (GAI)}
\label{subsec:alpha_investing}


The first proposals for online error rate control were based on ``alpha-investing'' by~\cite{Foster2008} and its generalisation (GAI) \citep{Aharoni2014}. {(An alternative early line of work instead focused on extensions of gatekeeping procedures that allow for online control of the FWER or FDR for ordered hypotheses~\citep{finos2011k, farcomeni2013fdr} but these turn out to be far less powerful in practice, so we do not discuss them further.)} Any GAI rule begins with an error budget, or \textit{alpha-wealth}, which is allocated to the different hypothesis tests over time. That is, there is a price to be paid each time a hypothesis is tested, {which can be viewed as making an investment in the hypothesis in question}. {If the} hypothesis is rejected, alpha-wealth is earned back, which can be viewed as a return {or payout} on the alpha-investment. {Since the alpha-wealth can increase in this way,} as long as discoveries continue to be made, hypotheses can be tested indefinitely without the test levels tending towards zero. {The intuition behind the alpha-wealth increasing after a rejection is that the denominator in the FDP increases, therefore allowing the numerator (i.e.\ the number of false rejections) to also increase for future hypothesis tests while still controlling the FDR}.

\begin{figure*}[ht!]
\includegraphics[width=\linewidth]{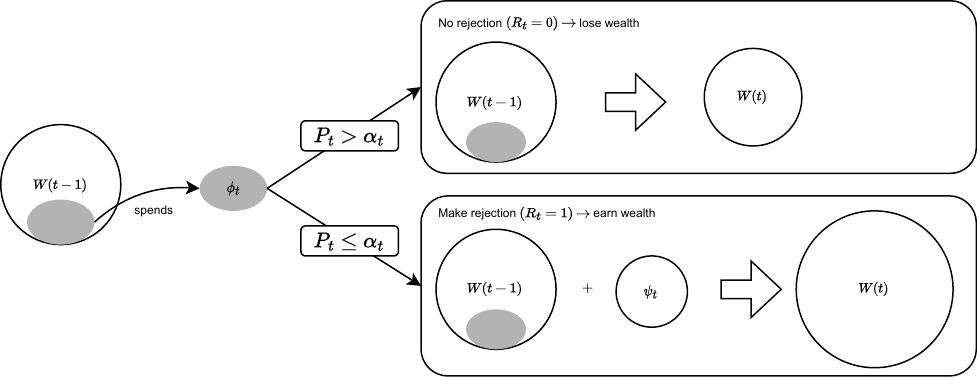}
\caption{{Diagrammatic representation of generalised alpha-investing (GAI), showing how the wealth $W(t)$ at time $t$ changes depending on whether the hypothesis $H_t$ is rejected (i.e.\ whether the corresponding $p$-value $P_t \leq \alpha$) or not. Figure adapted from~\citet{xu2022dynamic}.}}
\label{fig:alpha_investing}
\end{figure*}

Formally, a GAI rule produces a series of test levels $(\alpha_1, \alpha_2, \alpha_3, \ldots)$ based on which it uses~\eqref{eq:general-rejection} to produce the corresponding decisions $(R_1, R_2, R_3, \ldots)$. Of course, $\alpha_t$ must be based only on $R_1,\dots,R_{t-1}$. At each time point~$t$, {the} alpha-wealth $W(t)$ {decreases by an amount $\phi_t$}. If {the} hypothesis $H_t$ is rejected ($R_t=1$), then the alpha-wealth is increased by $\psi_t$.
In other words, the {price} $\phi_t$ is the amount paid for testing {(i.e., investing in)} a new hypothesis, and the payout (or {return on the investment}) $\psi_t$ is the amount earned if a discovery is made at that time.
Hence the initial wealth is $W(0) = w_0$ and it is updated via: \begin{align*}
W(t) & = W(t-1) - \phi_t + R_t \psi_t.
\label{eq:wealth_update}
\end{align*}
{Figure~\ref{fig:alpha_investing} give a diagrammatic summary of how GAI works.} The total wealth $W(t)$ must always be non-negative, and hence $\phi_t \leq W(t-1)$. Additionally, there are restrictions on $\alpha_t, \phi_t, \psi_t$, namely that when a rejection is made, the payout $\psi_t$ is capped. {This upper bound is there to ensure control of the FDR (and its variants)}.

{Given these constraints, the user is free to choose} the sequences $\alpha_t, \phi_t$ and $\psi_t$. {As an example, the alpha-investing rule explored in~\citet{Foster2008} chooses \[\alpha_t = \frac{\phi_t}{1+\phi_t}, \, \phi_t \leq W(t-1)\] and \[\psi_t = \phi_t + \alpha. \] The choice of $\alpha_t, \phi_t$ and $\psi_t$ was explored in terms of the trade-off between the sequences $\alpha_t$ and $\psi_t$ in \citet{Aharoni2014}. However, in this paper, we focus on the new `statistical' perspective for constructing online algorithms that control the FDR (see the start of Section~\ref{subsec:algorithms}), which \textit{implicitly} give choices for $\phi_t$ and $\psi_t$}. As is predominately the case in offline multiple testing, we often use \textit{monotone} decision rules for $\alpha_t$ considered as a function of $(R_1, \ldots, R_t)$: if $\tilde{R}_i \geq R_i$ for all $i \leq t-1$,  then we have $\alpha_t(\tilde{R}_1, \ldots, \tilde{R}_{t-1}) \geq \alpha_t(R_1, \ldots, R_t)$.

\citet{Ramdas2017} defined a class of improved GAI algorithms, called GAI++, as follows. Set $w_0$ so that $0 \leq w_0 \leq \alpha$ and choose the {payout} $\psi_t$ to satisfy \[\psi_t \leq \min\{ \phi_t + b_t, \frac{\phi_t}{\alpha_t} + b_t - 1\}, \] where {$b_t = \alpha - w_0 1\{R(t-1) = 0\}$. This upper bound on the payout is different from the original GAI algorithms in order to guarantee FDR control while giving the largest possible payout for rejecting a hypothesis, with the choice of $w_0$ determining the payout received for the very first rejection (see e.g.\ the LORD++ algorithm in Section~\ref{subsec:algorithms}). \citet{Ramdas2017} show that any monotone GAI++ rule comes with the following guarantee:}
\begin{theorem}
If the null $p$-values {(i.e., the subsequence of $p$-values where the null hypothesis is true)} are independent of all other $p$-values, any monotone GAI++ rule satisfies the bound $\mathbb{E} \left[ \frac{V(t) + W(t)}{R(t) {\vee 1}} \right] \leq \alpha$ for all $t \geq 1$. Since $W(t) \geq 0$, the FDR is controlled at level $\alpha$. 
\end{theorem}
This is in contrast to the GAI rules (including alpha-investing as proposed by~\citet{Foster2008}), which only control the mFDR.

{We note that the independence assumption refers to independence between \textit{different} hypotheses. The important case of sequential testing of any \textit{single} hypothesis can be seamlessly incorporated through the use of anytime-valid $p$-values as described in the Introduction. A related framework is to use `asynchronous' online testing, as discussed in Section~\ref{sec:extensions}, which gives the added flexibility of allowing hypothesis tests to overlap in time.}

\subsection{Algorithms for online FDR control: LORD, SAFFRON and ADDIS}
\label{subsec:algorithms}

Although an ``algorithmic perspective'' led to the GAI procedures initially used in the online testing literature,~\cite{Ramdas2017} posited a ``statistical perspective'' to construct procedures, which is to keep an estimate of the FDP less than $\alpha$. First, the oracle FDP is defined as \[
\text{FDP}^*(t) = \frac{\sum_{j \leq t, j \in \mathcal{H}_0} \alpha_j}{R(t) \vee 1},
\]
where $\mathcal{H}_0$ denotes the set of true null hypotheses. If we can keep $\text{FDP}^*(t) \leq \alpha$ at all times~$t$, then (depending on dependence assumption on the $p$-values) we can prove that $\text{mFDR}(t) \leq \alpha$ or $\text{FDR}(t) \leq \alpha$. This technique has been used to derive the LORD, SAFFRON and ADDIS algorithms (see below), by designing different estimates $\widehat{\text{FDP}}_{\text{LORD}}(t), \widehat{\text{FDP}}_{\text{SAFFRON}}(t), \widehat{\text{FDP}}_{\text{ADDIS}}(t)$ for $\text{FDP}^*(t)$. 

\subsubsection{LORD}
The LORD algorithm was conceptualised by ~\citet{Javanmard2018}, and is an instance of a monotone GAI rule. More precisely, given an infinite non-increasing sequence of positive constants $\{ \gamma_t\}_{t=1}^{\infty}$ that sums to one, the test levels $\alpha_t$ for LORD are chosen as follows:
\begin{equation*}
\alpha_t =  w_0\gamma_t + \sum_{j \, : \, \tau_j < t, \, \tau_j \neq \tau_1} \gamma_{t - \tau_j} b_0,
\label{eq:LORD}
\end{equation*}
where $\tau_j$ denotes the time of the $j$-th rejection and we must have $w_0 + b_0 \leq \alpha$ for FDR control to hold.

Following this, \citet{Ramdas2017} defined a simple upper bound of $\text{FDP}^*(t)$: \[
\widehat{\text{FDP}}_{\text{LORD}}(t) = {\frac{\hat{V}(t)}{R(t) \vee 1}} := \frac{\sum_{j \leq t} \alpha_{{j}}}{R(t) \vee 1}
\]
and showed that LORD can be viewed as an algorithm that keeps $\widehat{\text{FDP}}_{\text{LORD}}(t)\leq \alpha$. {Here $\hat{V}(t)$ corresponds to the alpha-wealth used for testing while $\alpha R(t)$ corresponds to the total \textit{earned} alpha-wealth that can be used for subsequent tests.}
Exploiting this view, they derived a uniform improvement of LORD, termed LORD++, (presented below). In brief, LORD++ is able to replace $b_0 = \alpha - w_0$ with the choice $b_0 = \alpha$ while still maintaining FDR control, with the catch that for the very first rejection only $b_0 = \alpha - w_0$ (see below).

Given an infinite non-increasing sequence of positive constants $\{ \gamma_t\}_{t=1}^{\infty}$ that sums to one, the test levels $\alpha_t$ for LORD++ are chosen as follows:
\begin{equation*}
\alpha_t =  w_0\gamma_t + (\alpha - w_0) \gamma_{t - \tau_1}\mathbbm{1}\{\tau_1<t\} + \alpha \sum_{j \, : \, \tau_j < t, \, \tau_j \neq \tau_1} \gamma_{t - \tau_j}.
\label{eq:LORD++}
\end{equation*}
The above formula may look daunting but it is interpretable. The first term is the fraction of the initial wealth $w_0$ that is used by the $t$-th test. The other terms are the fractions of the earnings from rejections before $t$ that are spent in the $t$-th round: LORD++ awards $\alpha-w_0$ for the first rejection and $\alpha$ for every subsequent rejection, and on receiving this reward, the method immediately allocates that reward to future rounds according to the same schedule of constants $\{\gamma_t\}$, shifted to start at the next instant. This rule ensures that LORD++ never spends more than it has earned, thus keeping $\widehat{\text{FDP}}_{\text{LORD++}}(t)\leq \alpha$.

The intuitive reason why LORD++ cannot award $\alpha$ for the very first rejection can be seen in the definition $\text{FDP}(T) = \frac{V(T)}{R(T) \vee 1}$. The denominator $R(T) \vee 1 = 1$ when the number of rejections equals zero or one, and hence only starts increasing at the second rejection. This means that the sum of $w_0$ and the first reward must be at most~$\alpha$, following which $\alpha$ may be rewarded at every rejection. As for the choice of the sequence $\gamma_t$, this depends on the data application at hand, with a reasonable default choice given by $\gamma_t \propto \frac{\log (t \vee 2)}{t \, \text{exp}(\sqrt{\log t})}$, which has been shown to maximise power in the Gaussian setting (i.e.\ where the test statistics follow a normal distribution)~\citep{Javanmard2018}. 

The manner in which $\widehat{\text{FDP}}_{\text{LORD++}}(t)$ is a simple upper bound on $\text{FDP}^*(t)$ is reminiscent of the BH procedure for offline testing, which can be derived in a similar fashion. More precisely, suppose that one rejects all $p$-values below some fixed threshold $s \in (0,1)$. The BH 
procedure overestimates the FDP using the quantity $\widehat{\text{FDP}}_{\text{BH}}(s) = \frac{n \cdot s}{|\mathcal{R}(s)|}$, where $\mathcal{R}(s)$ denotes the set of rejected $p$-values using the fixed threshold~$s$. The BH procedure then rejects the set $\mathcal{R}(\hat{s}_{\text{BH}})$ where $\hat{s}_{\text{BH}} = \max\{s : \widehat{\text{FDP}}_{\text{BH}}(s) \leq \alpha\}$. This leads us to view LORD++ as the online analog of the BH procedure.

Guarantees for LORD++ hold under different $p$-value dependencies, which we now formalise. 
Define the filtration at time~$t$ as $\mathcal{F}^t = \sigma(R_1, \ldots, R_t)$ (representing the collection of the observed rejections up to time~$t$)
and let $\alpha_t = f_t(R_1, \ldots, R_{t-1})$ where $f_t$ is a $[0,1]$-valued function. 
The null $p$-values are said to be conditionally super-uniform if $\text{Pr}\{P_t \leq \alpha_t | \mathcal{F}^{t-1}\} \leq \alpha_t$ for any $\mathcal{F}^{t-1}$-measurable $\alpha_t$. 
Armed with this definition, we have the following theorem from~\cite{Ramdas2017}:
\begin{theorem}
(a) If the null $p$-values are conditionally super-uniform, then the condition $\widehat{\text{FDP}}_{\text{LORD}}(t) \leq \alpha$ for all $t \geq 1$ implies that $\text{mFDR}(t) \leq \alpha$ for all $t \geq 1$. \\
(b) If the null $p$-values are independent of each other and of the {$p$-values corresponding to the} non-null {hypotheses}, and $\{\alpha_t\}$ {is chosen} to be a monotone function of past rejections, then the condition $\widehat{\text{FDP}}_{\text{LORD}}(t) \leq \alpha$ for all $t \geq 1$ implies that $\text{FDR}(t) \leq \alpha$ for all $t \geq 1$.
\end{theorem}

Finally, in terms of theoretical power guarantees for LORD, \citet{chen2021power} considered the setting of a (generalised) Gaussian model (see reference for further details) and showed that LORD is asymptotically optimal, in particular by being as powerful as BH to first asymptotic order.

\subsubsection{SAFFRON}
\citet{Ramdas2018} derived an adaptive version of LORD++ called SAFFRON, 
which is based on an estimate of the proportion of true null hypotheses.
By not wasting its earnings on attempting to reject weaker signals (i.e.\ larger $p$-values), SAFFRON preserves alpha-wealth and hence can have a higher power than LORD++. 
To this end, {we} choose $\lambda \in (0,1)$ and define the candidate $p$-values as those that satisfy $P_t \leq \lambda$, since SAFFRON will never reject a $p$-value larger than~$\lambda$. 
{We also choose an infinite nonincreasing sequence of positive constants $\{ \gamma_t\}_{t=1}^{\infty}$ that sums to one. Reasonable default choices for these hyper-parameters are $\lambda = 0.5$ and $\gamma_t \propto t^{-1.6}$~\citep{Ramdas2018}. The formulae for the test levels~$\alpha_t$ for SAFFRON are given in Appendix~\ref{Asec:SAFFRON}}. 

SAFFRON starts off with alpha-wealth $(1 - \lambda) w_0$ and does not lose any of this wealth when testing candidate $p$-values. {Of course, this has to be done in a principled way and is accounted for in the formulation of the test levels $\alpha_t$, which intuitively helps explain the $(1-\lambda)$ multiplicative factor (see Appendix~\ref{Asec:SAFFRON})}. It gains an alpha-wealth of $(1-\lambda)\alpha$ for each discovery after the first. SAFFRON can make more rejections than LORD++ if there is a significant fraction of non-nulls and the signals are strong.

Similar to LORD++, SAFFRON provably controls the mFDR at all times if the null $p$-values are conditionally super-uniform. Also, SAFFRON controls the FDR at all times if the null $p$-values are independent of each other and of the non-nulls, and $\{\alpha_t\}$ is chosen to be a monotone function of $(R_1, \ldots, R_{t-1}, C_1, \ldots, C_{t-1})$, {where $C_t= \mathbbm{1}\{ P_t \leq \lambda\}$}; see~\citet{Ramdas2018} for details.

\subsubsection{ADDIS} stands for an ADaptive algorithm that DIScards conservative nulls, and was proposed by~\citet{Tian2019}. ADDIS can invest alpha-wealth more effectively than LORD++ or SAFFRON by explicitly discarding the weakest signals (i.e.\ the largest $p$-values) in a principled way, which can lead to a higher power. More formally, in practice it is common to encounter \textit{conservative} nulls, where a null $p$-value $P$ is conservative if $\text{Pr}\{P \leq x\} < x$ for all $x \in [0,1]$. Often nulls are \textit{uniformly} conservative, which means that under the null, 
\[
\text{Pr}\{P/c \leq x \, | \, P \leq c \} \leq x \quad \text{for all } x, c \in (0,1).
\]
For example, for a one-dimensional exponential family with parameter~$\theta$, when the true parameter~$\theta$ is strictly smaller than $\theta_0$, the uniformly most powerful test of $H_0: \theta \leq \theta_0$ versus $H_1 : \theta > \theta_0$ will give uniformly conservative nulls~\citep{zhao2019multiple}. Another setting is using always-valid $p$-values~\citep{johari2021always} in the context of continuous monitoring for A/B testing, which will always be conservative.

In general, adaptivity (used by both SAFFRON and ADDIS) helps when there is a significant fraction of non-nulls (like 10\% or 20\%). Discarding (used only by ADDIS) helps when the nulls are conservative, meaning that instead of being exactly uniform, they are stochastically larger than uniform. Discarding helps even without adaptivity, and adaptivity helps without discarding. The key idea behind discarding is intuitive: if you see a $p$-value larger than (say) 0.5, throw it away, but if you see a $p$-value smaller than 0.5, then double it (to condition on selection) and pass it onto the multiple testing procedure. Roughly, if there are mostly nulls and these are uniformly distributed, this doesn't do much at all -- the tested $p$-values are doubled, but only about half the $p$-values are tested so the multiplicity correction is halved, cancelling the effects. However, if the nulls are stochastically much larger than uniform, then we may throw away most of the nulls in this step, eventually testing only a much smaller number of $p$-values (which have been doubled).

In terms of formal definitions, with $\lambda$ and the corresponding candidate $p$-values defined as for SAFFRON, we let $S_t = \mathbbm{1} \{ P_t \leq \eta \}$ be the indicator of $H_t$ being selected for testing (i.e.\ not discarded). Hence $\eta$ is the discarding threshold and must be greater than $\lambda$. We also choose an infinite non-increasing sequence of positive constants $\{ \gamma_t\}_{t=0}^{\infty}$ that sums to one. {Reasonable default choices for these hyper-parameters are $\lambda = 0.25, \eta = 0.5$ and $\gamma_t \propto (t+1)^{-1.6}$, as justified empirically in~\citet{Tian2019}. The formulae for the test levels~$\alpha_t$ for ADDIS are given in Appendix~\ref{Asec:ADDIS}}. As can be seen, ADDIS starts off with an alpha-wealth of $(\eta - \lambda)w_0$ and (like SAFFRON) does not lose any of this wealth when testing candidate $p$-values. The $p$-values that are greater than $\eta$ do not affect the test levels for ADDIS at all, i.e.\ as if they did not exist in the sequence of $p$-values of all (reflecting the term `discarding'). It gains an alpha-wealth of $(\eta - \lambda)\alpha$ for each rejection after the first. 

Like for LORD++ and SAFFRON, ADDIS provably controls the mFDR at all times if the null $p$-values are conditionally uniformly conservative. ADDIS provably controls the FDR at all times if the null $p$-values are independent of each other and of the non-nulls, and $\{\alpha_t\}_{t=1}^{\infty}$ is a monotone function of the past; see~\citet{Tian2019} for full details.

\subsubsection{Monotone AI} As a comparator to the above algorithms, we also consider a version of the original AI algorithm of~\cite{Foster2008}, as modified by~\citet{Ramdas2017} to ensure it is a monotone rule and hence that FDR control holds. We will refer to this rule as `monotone AI'. 


\section{Simulation studies}
\label{sec:simulation}

In this section we compare the performance of the LORD++, SAFFRON and ADDIS algorithms in terms of the FDR and statistical power.
We do not aim to present an exhaustive simulation of all the algorithms currently available in the literature, but rather select a representative set of algorithms to demonstrate some key general features for the core problem of online FDR control. To this end, we use LORD++ as a representative ‘basic’ online algorithm, given that it is the natural online analog of the BH procedure. We then use SAFFRON as a representative of an adaptive online algorithm, while ADDIS is a representative of an adaptive online algorithm that also incorporates discarding. As an additional comparison, we also include the `monotone AI' rule. 
In Section~\ref{subsec:further_sim} we refer the reader to further simulation {studies} that have been published in the literature. First though, we briefly describe software implementation of algorithms for online error rate control.

\subsection{Software: \texttt{onlineFDR} package}
\label{subsec:software}

The \texttt{onlineFDR} package is an open-source R package that aims to provide a comprehensive and up-to-date implementation of algorithms for online error rate control. It is freely available via \textit{Bioconductor}~\citep{onlineFDR}. 
The package implements the LORD++, SAFFRON and ADDIS algorithms, as well as almost all of the algorithms corresponding to the further extensions of online error rate control methodology (see Section~\ref{subsec:further_extensions}). In particular, it also provides functions for algorithms for online FWER and online FDX control.
The package documentation provides a user-friendly introduction to the use of the package, and there is also a Shiny app available~\citep{onlineFDR_shiny} to allow users to explore algorithms for online FDR control in an interactive way without having to program. All results for the simulation and case studies in this paper were calculated using the package.

\subsection{Testing with Gaussian observations}
\label{subsec:test_gaussian}

In order to examine the relative performance of the online FDR algorithms, we use a simple experimental setup of testing Gaussian means, with a total of~$T$ hypotheses. Note that although~$T$ is fixed in the simulations, all the methods do not use this knowledge of~$T$ to normalise the sequence $\gamma_t$. The null hypotheses take the form $H_t : \mu_t \leq 0$ which are tested against the alternative $H_t' : \mu_t > 0$ for $t = 1, \ldots, T$. We observe independent observations $Z_t \sim N(\mu_t, 1)$ which are transformed to one-sided $p$-values $P_t = \Phi(-Z_t)$, where $\Phi$ denotes the standard Gaussian CDF. {The motivation for using one-sided $p$-values is from A/B testing, where one wishes to detect larger effects, not smaller}. The means $\mu_t$ are set according to the following mixture model: 
\[
\mu_t = \begin{cases}
F_0 \qquad \text{with probability } 1 - \pi_1 \\
F_1 \qquad \text{with probability } \pi_1
\end{cases} ,
\] 
where $F_1 \sim N(3, 1)$ and $F_0$ is defined as below.

We use the default settings for LORD++, SAFFRON and ADDIS that are implemented in the \texttt{onlineFDR} package (following suggestions in the literature). For LORD++, we {use} the {default choice of} $\gamma_t \propto \frac{\log (t \vee 2)}{t \, \text{exp}(\sqrt{\log t})}$. We also use this choice of $\gamma_t$ for alpha-spending, where $\{ \alpha_t \}_{t=1}^{\infty}$ is simply given by $\alpha_t = \alpha \gamma_t$. For SAFFRON we set $\lambda = 0.5$ and $\gamma_t \propto \frac{1}{t^{1.6}}$. Finally, for ADDIS we set $\lambda = 0.25$, $\eta = 0.5$ and $\gamma_t \propto \frac{1}{(t+1)^{1.6}}$. {We use the exponent $1.6$ in the denominator for $\gamma_t$ because this was found empirically to work well in a range of different simulation studies in the original papers. More precisely, the sequence $\gamma_t$ satisfies $\sum_{t=1}^{\infty} \gamma_t = 1$.}

\begin{figure}[ht!]
\centering
\includegraphics[width=0.85\linewidth]{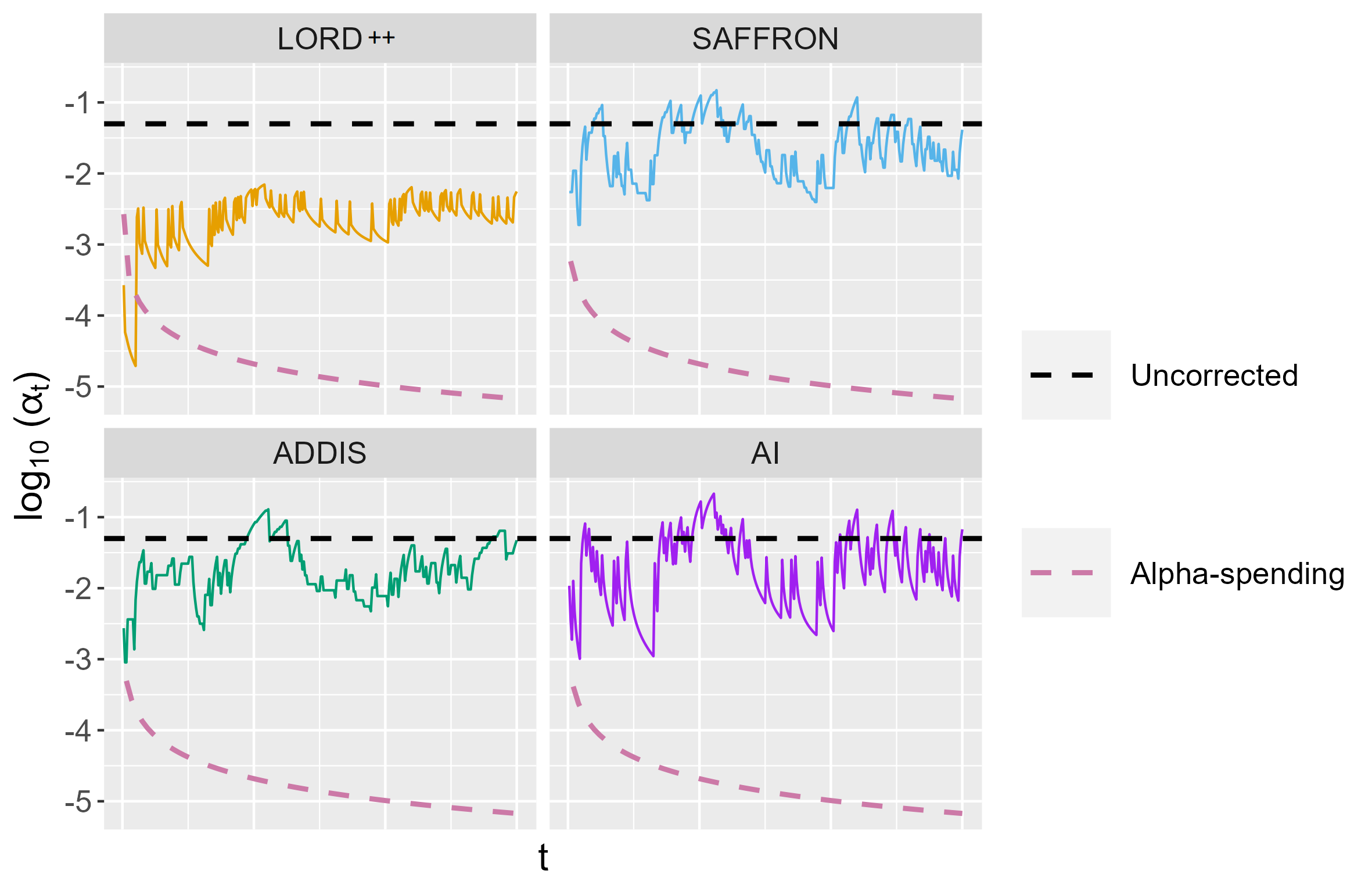}
\caption{Test levels for LORD++, SAFFRON, ADDIS {and monotone AI} compared with uncorrected testing and alpha-spending. We set $T = 300$, $\alpha = 0.05$ and the proportion of non-nulls $\pi_1 = 0.5$.}
\label{fig:test_levels}
\end{figure}

Figure~\ref{fig:test_levels} shows how the test levels $\{ \alpha_t \}_{t=1}^{\infty}$ (displayed on the $\log_{10}$ scale) evolve over time for LORD++, SAFFRON, ADDIS {and monotone AI} compared with uncorrected testing (where $\alpha_t \equiv \alpha$) and alpha-spending. Here, $T = 300$, $\alpha = 0.05$, $\pi_1 = 0.5$ and we choose $F_0 \equiv 0$. 

All of the online FDR algorithms have higher test levels than alpha-spending (apart from LORD++ briefly early on in this particular experiment). The relative difference increases with~$t$ as the online algorithms `earn back' wealth over time, which alpha-spending cannot do. SAFFRON, ADDIS {and monotone AI} have higher test levels than LORD++, reflecting how they can more efficiently invest the alpha-wealth. In this setting, since $\mu_t = 0$ under the null, the nulls are exactly uniform and so ADDIS cannot take advantage of conservative nulls. Hence the testing levels of ADDIS are similar or slightly lower than those for SAFFRON {and monotone AI}. Finally, we see that SAFFRON has similar test levels as uncorrected testing, and SAFFRON, ADDIS {and monotone AI} can even have test levels above the nominal $\alpha$.

\begin{figure}[ht!]
\centering
\includegraphics[width=0.95\linewidth]{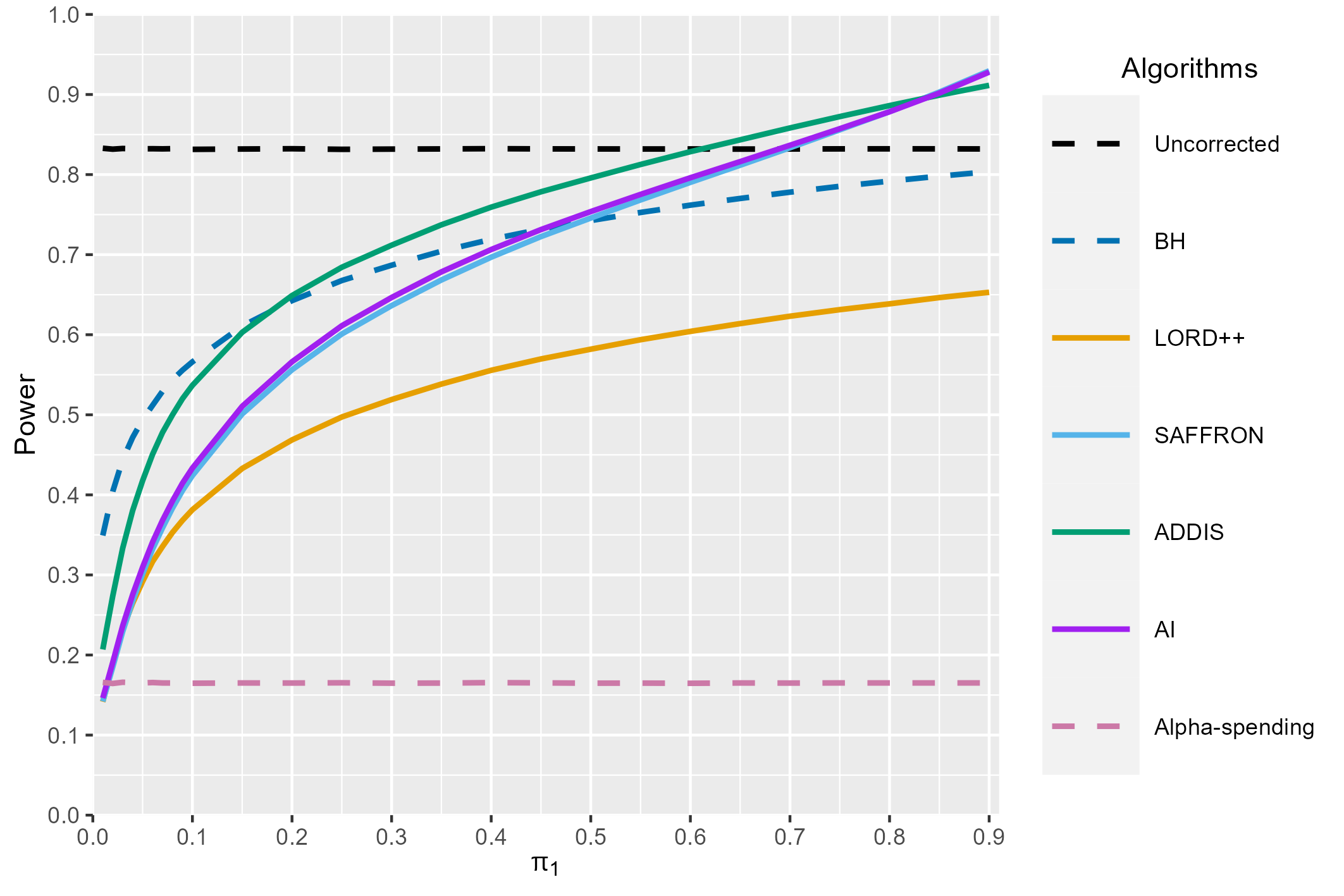}
\caption{Power of LORD++, SAFFRON, ADDIS {and monotone AI} compared with uncorrected testing, the BH procedure and alpha-spending as the proportion of non-nulls $\pi_1$ varies. We set $T = 1000$ and $\alpha = 0.05$. Results are based on $10^4$ simulation replicates.}
\label{fig:power}
\end{figure}

Figure~\ref{fig:power} compares the statistical power of LORD++, SAFFRON, ADDIS {and monotone AI} compared with uncorrected testing and alpha-spending, as $\pi_1$ varies from 0.01 to 0.9. Here, we define power as \[
\text{power}(T) = \mathbb{E} \left[ \frac{\sum_{t \in \mathcal{H}_1} R_t}{\left(\sum_{t=1}^T \mathbbm{1}\{t \in \mathcal{H}_1\}\right) \vee 1} \right] ,
\]
where $\mathcal{H}_1$ denotes the index set of the non-null hypotheses. We also include the standard Benjamini-Hochberg (BH) procedure as an additional comparison. We stress that BH is an offline procedure and so could not be used for online testing in practice. In our simulation, we set $T = 1000$, $\alpha = 0.05$ and $F_0 \sim N(-0.5, 0.1)$. Results are based on averaging $10^4$ simulation replicates (which implies a Monte Carol standard error when power = 0.5 of 0.005).

Starting with alpha-spending, as expected the power is very low ($<0.2$) for all $\pi_1$. LORD++ has substantial power gains compared with alpha-spending (as long as $\pi_1$ is not close to zero) and this advantage increases with~$\pi_1$. However, LORD++ has substantially lower power than BH for all values of $\pi_1$. As expected, SAFFRON performs better as the fraction of non-nulls $\pi_1$ increases, with a higher power than LORD++ for $\pi_1 > 0.05$, BH for $\pi_1 > 0.5$ and even uncorrected testing for $\pi_1 > 0.7$. Since $F_0 \sim N(-0.5, 0.1)$, almost all the means for the null hypotheses will be negative, i.e.\ we are in a setting with conservative nulls. Hence, as expected, ADDIS outperforms SAFFRON in terms of power (except for very high values of $\pi_1$). ADDIS also has a higher power than BH for $\pi_1 > 0.2$ and uncorrected testing for $\pi_1 > 0.6$. {Finally, in this setting, SAFFRON performs very similarly to the monotone AI algorithm in terms of power.}

In Appendix~\ref{Asec:sim} (Figure~\ref{fig:FDR}) we show the corresponding FDR for all of the algorithms considered. We see that uncorrected testing can have substantial inflation of the FDR, with the FDR inflated above the nominal $\alpha = 0.05$ level for $\pi_1 < 0.3$. The FDR reaches as high as 0.65 when $\pi_1 = 0.01$. All other algorithms control the FDR below the nominal 0.05 level, as expected.

\subsection{Observations from other simulations}
\label{subsec:further_sim}

Here, we summarise a few take-home messages for LORD, SAFFRON and ADDIS from simulation results already found in the literature. \citet{Javanmard2018} investigated the effect of the ordering of the hypotheses for online testing rules, including LORD. In some applications, hypotheses can be ordered using side information, such that those that are most likely to be rejected come first. With this favourable ordering, the statistical power of LORD can substantially increase as long as $\pi_1$ is not too large (since ordering is less relevant in that case). Similar findings for LORD++, SAFFRON and ADDIS in the context of platform trials can be found in~\citet{robertson2022online}, which also looked at the adversarial setting where hypotheses happen to be ordered so that those most likely to be rejected come last, resulting in lower power.

\citet{Ramdas2018} considered the impact on LORD++ and SAFFRON of choosing sequences of the form $\gamma_t \propto t^{-s}$, where the parameter $s > 1$ controls the `aggressiveness' of the procedure (since the larger the value of~$s$, the more the alpha-wealth is concentrated at small values of~$t$). For Gaussian alternatives, the simulation results suggested that less aggressive sequences are to be preferred in terms of increased power for SAFFRON and LORD++. 
Meanwhile, \citet{Tian2019} showed that ADDIS can match the power of SAFFRON when the nulls are not conservative (i.e.\ uniform nulls). The power advantage of ADDIS over LORD++ and SAFFRON increases the more conservative the nulls are, i.e.\ the more negative the means for the null hypotheses are (in the Gaussian setting).

The theory presented in Section~\ref{sec:methods} for provable FDR control requires null $p$-values to be independent of one another (with a weaker condition sufficing for mFDR control). {\citet{robertson2022online} explored the performance of online testing rules, including LORD++, SAFFRON and ADDIS, in the setting of platform trials with a common control, which induces positive correlations between the $p$-values for testing concurrent arms. There was no evidence of FDR inflation for these algorithms under a range of assumed treatment effects and overlap of control data}. \citet{Robertson2018} {considered the setting where} the test statistics are assumed to come from a multivariate normal distribution where the covariance matrix has ones along the diagonals and off-diagonal entries equal to $\pm 0.5$. There was no evidence of FDR inflation when using LORD++ under a range of non-null distributions. However, with a two-sided test under a Gaussian alternative, the SAFFRON procedure had an inflated FDR for smaller values of $\pi_1$. This inflation persisted and even increased as~$T$ increased from 100 to 1000. {For further discussion handling dependent $p$-values, we refer the reader to the end of Section~\ref{subsec:future}}.

The simulation studies in~\citet{Robertson2018} also highlighted the value of using `bounded' versions of online testing algorithms. This requires setting an \textit{a-priori} upper bound~$M$ on the number of hypotheses to be tested, so that the $\gamma_t \equiv 0$ for $t>M$, which allows setting $\gamma_t \equiv 1/M$ for $t \leq M$ (for example). The bounded versions have a uniformly higher power than the versions presented in Sections~\ref{sec:methods} (with the default choices of $\gamma_t$ given in Section~\ref{subsec:test_gaussian}) which assume no upper bound on~$T$, and empirically a substantial gain can be observed for small~$T$ (i.e.\ $T < 100$). Finally, another general observation is that the power advantages of online testing algorithms compared with alpha-spending increase as~$T$ increases. Indeed, when~$T$ is small and $\pi_1$ is low, online testing algorithms may no longer be competitive in terms of power. We return to this issue in Section~\ref{subsec:future}.


\section{Case studies}
\label{sec:case_studies}

\subsection{IMPC dataset}
\label{subsec:IMPC}

Our first case study uses high-throughput phenotypic data from the International Mouse Phenotyping Consortium (IMPC) data repository, which aims to generate and phenotypically characterize knockout mutant strains for every protein-coding gene in the mouse~\citep{Koscielny2013}. The IMPC database is an example of a growing dataset mentioned in Section~\ref{subsec:motivation}, since the family of hypotheses is constantly growing as new knockout mice lines are generated and phenotyping data is uploaded to the data repository.

We focus on the analysis of IMPC data performed by \citet{Karp2017}, who looked at the influence of sex in mammalian phenotypic traits in both wildtype and mutants. As part of their analysis, Karp et al.\ analysed the role of sex as a modifier of the genotype effect (for continuous traits) using a two stage pipeline. Stage~1 tested the role of genotype using a likelihood ratio test comparing models~(a) and~(c). Similarly, stage~2 tested the role of sex using a likelihood ratio test comparing models~(a) and~(b). 
\small
\begin{align*}
(a) \;  Y & \sim \text{Genotype} + \text{Sex} + \text{Genotype}*\text{Sex} + \text{Weight} + (1|\text{Batch})  \\
(b) \; Y & \sim \text{Genotype} + \text{Sex} + \text{Weight} + (1|\text{Batch}) \\
(c) \; Y & \sim \text{Sex} + \text{Weight} + (1|\text{Batch})
\end{align*}
\normalsize

The above procedure resulted in two sets of $N = 172\,328$ distinct $p$-values, ordered by the date of the corresponding genomic assay. Note that these $p$-values will not be independent, due to positive and negative associations between different genes (caused for instance by linkage disequilibrium). In addition, multiple variables are being measured for the same gene, and these can be aspects of the same phenotype or be biologically correlated.

Table~\ref{tab:IMPC} below shows the number of traits that had a statistically significant genotype effect or were classed as having a statistically significant sexual dimorphism (SD) using LORD++, SAFFRON, ADDIS {and monotone AI}. As a comparison, we include the results from alpha-spending, BH and uncorrected testing. The `Fixed Threshold' procedure is the fixed significance threshold of 0.0001 used in practice for the IMPC pipeline.

\begin{table}[ht!]
	\centering
	
	\begin{tabular}{p{3cm} p{2cm} p{2cm}} 
	 & \textbf{Genotype} & \textbf{SD} \\ \hline
	\textbf{Uncorrected} & 35\,575 & 20\,887  \\
    \textbf{BH} & 12\,907 & 2\,084  \\
	\textbf{ADDIS} & 12\,558  &  1\,713 \\
	\textbf{SAFFRON} & 14\,268 & 1\,705  \\
	\textbf{LORD++} & 8\,517 & 1\,193  \\
    \textbf{{Monotone AI}} & {9\,906} & {985} \\ 
	\textbf{Fixed Threshold} & 4\,158 & 969 \\
	\textbf{Alpha-spending} & 795 &  60 \\ \hline
	\end{tabular}
	
	\caption{Number of rejections made by online FDR algorithms and various comparators using the IMPC datasets. SD = Sexual Dimorphism. \label{tab:IMPC}}
\end{table}

Starting first with the results for the genotype data, the online testing algorithms make two to three times as many rejections as fixed testing. ADDIS and SAFFRON in turn make substantially more rejections than LORD++, with an increase of almost 50\% and 70\%, respectively. ADDIS makes a similar number of rejections to BH, but SAFFRON makes noticeably more rejections than both BH and ADDIS for these data. For the SD data, again the online testing algorithms make substantially more rejections than fixed testing, but the relative increase is much less. ADDIS and SAFFRON make a very similar number of rejections, about 50\% more than the number of rejections for LORD++. {Finally, for these data we see that monotone AI makes substantially fewer rejections then either ADDIS or SAFFRON.}

\subsection{Platform trial: STAMPEDE}
\label{subsec:STAMPEDE}

\begin{figure}[ht!]
\centering
\includegraphics[width=0.9\linewidth]{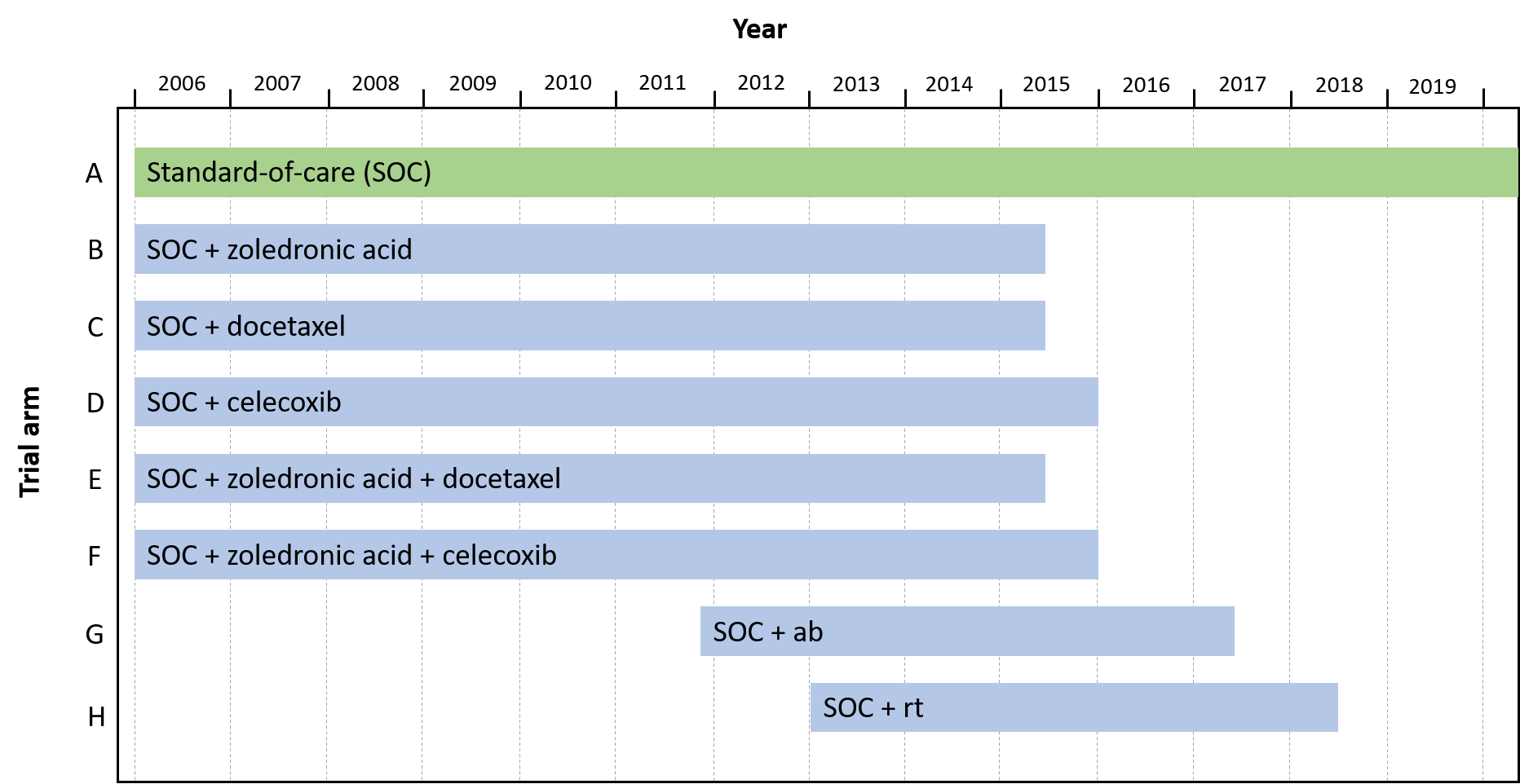}
\caption{Schematic of the completed treatment arms in the STAMPEDE platform trial. ab = abiraterone, rt = radiotherapy.}
\label{fig:STAMPEDE}
\end{figure}

Our second case study is the ongoing STAMPEDE (Systemic Therapy for Advancing or Metastatic Prostate Cancer) platform trial, which evaluates the effect of systemic therapies for prostate cancer on overall survival~\citep{james2008stampede}. The trial started with 5 experimental treatment arms (B-–F), and compared these with the control arm A, which was standard-of-care (SOC) hormone therapy. Figure~\ref{fig:STAMPEDE} shows a schematic of the treatment comparisons that have already been reported from STAMPEDE. Two additional experimental arms (G and H) were added to the trial in 2011 and 2013, respectively.

Table~\ref{tab:STAMPEDE_pval} shows the reported $p$-values (unadjusted for multiplicity) when comparing the experimental arms with the control (arm A), as given in \citet{james2016stampede, james2017stampede, mason2017stampede, parker2018stampede}. The dashed lines denote the four `batches' present in the trial, where a batch corresponds to multiple hypotheses being available to be tested at the same time, as reflected in Figure~\ref{fig:STAMPEDE}.

\begin{table}[ht!]
    \centering
    \begin{tabular}{l l}
         \textbf{Trial arm} & \textbf{$p$-value} \\ \hline
         \textbf{B}: SOC + zoledronic acid & 0.450 \\
         \textbf{C}: SOC + docetaxel & 0.006 \\
         \textbf{E}: SOC + zoledronic acid + docetaxel & 0.022 \\ \hdashline
         \textbf{D}: SOC + celecoxib & 0.847 \\
         \textbf{F}: SOC + zoledronic acid + celecoxib & 0.130 \\ \hdashline
         \textbf{G}: SOC + abiraterone & 0.001 \\ \hdashline
         \textbf{H}: SOC + radiotherapy & 0.266 \\[6pt]
    \end{tabular}
    \caption{Reported $p$-values for the STAMPEDE platform trial. SOC = Standard-of-care.}
    \label{tab:STAMPEDE_pval}
\end{table}

Following~\citet{robertson2022online}, we apply the online testing algorithms to these observed $p$-values, keeping the alphabetical ordering of $p$-values within the batches. We set the upper bound on the number of treatments $M = 20$ (i.e.\ twice as many arms that have already entered the STAMPEDE trial as of the end of 2021), and use the bounded versions of alpha-spending (i.e.\ a Bonferroni correction at level $\alpha/M$), LORD++, SAFFRON and ADDIS. Table~\ref{tab:STAMPEDE_rej} shows which of the hypotheses corresponding to each experimental arm can be rejected at level $\alpha = 0.05$, as well as the current significance level $\alpha_8$ that would be used to test the next experimental treatment after the 7 already evaluated in the trial.

\begin{table}[ht!]
\centering
    \begin{tabular}{l | l | l}
         \textbf{Algorithm} & \textbf{Rejections} & \hspace{6pt}$\bm{\alpha_8}$ \\ \hline
         Uncorrected & C, E, G & 0.0500 \\
         Alpha-spending & G & 0.0025 \\
         BH & C, G & --\\
         ADDIS & G & 0.0016 \\
         SAFFRON & C, G & 0.0165\\
         LORD++ & -- & 0.0002\\
    \end{tabular}
    \vspace{6pt}
    \caption{Hypotheses rejected and current significance level $\alpha_8$ of different algorithms using the results of the STAMPEDE trial, with the ordering as in Table~\ref{tab:STAMPEDE_pval}.}
    \label{tab:STAMPEDE_rej}
\end{table}

Uncorrected testing rejects the hypotheses corresponding to three experimental arms (C, E, G), and has by far the highest value of~$\alpha_8$. Both SAFFRON and the BH procedure reject hypotheses~C and~G, and SAFFRON has a substantially higher value of~$\alpha_8$ than for the other online testing algorithms. ADDIS and alpha-spending only reject hypothesis~G, and have similar~$\alpha_8$. Finally, LORD++ does not reject any hypotheses and the value of~$\alpha_8$ is substantially lower than any of the other algorithms. For further discussion and results, see~\cite{robertson2022online}.


\section{Extensions and future directions}
\label{sec:extensions}

\subsection{Further extensions}
\label{subsec:further_extensions}

\subsubsection{Prior weights, penalty weights and decaying memory}
\citet{Ramdas2017} proposed a number of extensions that apply to the class of GAI++ algorithms, including LORD++. Firstly, they showed how to incorporate certain types of prior information about the different hypotheses as expressed through \textit{prior weights} $w_t$ and \textit{penalty weights} $u_t$. Prior weights allow the experimenter to exploit domain knowledge about which hypotheses are more likely to be non-null. By assigning a higher prior weight $w_t > 1$ to a hypothesis, the algorithm will have a higher chance of rejecting $H_t$. Meanwhile, penalty weights express the different importance attached to the hypotheses being tested, with $u_t > 1$ indicating a more impactful or important test. Importantly, both $w_t$ and $u_t$ are allowed to depend on past rejections in this framework. \citet{Ramdas2017} proposed doubly-weighted GAI++ rules that provably control the \textit{penalty-weighted} FDR when using both prior and penalty weights under independence. {Recently, \citet{pmlr-v108-chen20b} showed how to exploit contextual information associated with each hypothesis to re-weight the testing levels in an online manner, leading to increased power while controlling the FDR}.

The second proposal {of \citet{Ramdas2017}} dealt with problems of \textit{`piggybacking'} and \textit{`alpha-death'}. Piggybacking happens when a substantial number of rejections are made so that the online testing algorithms earn and accumulate enough alpha-wealth to reject later hypotheses at much less stringent thresholds (hence the later tests `piggyback' on the success of earlier tests). This can lead to a spike in the FDR \textit{locally in time}, even though the FDR over all time is controlled. Meanwhile, alpha-death occurs when there is a long stretch of null hypotheses, so that online testing algorithms make (almost) no rejections and lose nearly all of their alpha-wealth. Subsequently, the algorithm may have essentially no power, unless a non-null hypotheses with extremely strong signal (small $p$-value) is observed. \citet{Ramdas2017} proposed the \textit{decaying memory} FDR (mem-FDR), which pays more attention to recent discoveries through a user-defined discount factor $\delta \in (0,1]$ and thus smoothly forgets the past. They then proposed GAI++ rules that control the mem-FDR (which can also include penalty weights) under independence. In addition, they showed how to allow the algorithm to \textit{abstain} from testing in order to recover from alpha-death. \\

\subsubsection{Local dependence -- asynchronous and batched testing}
In most of the literature mentioned so far, an implicit assumption is that each hypothesis test can only start when the previous test has finished (the \textit{synchronous} setting, {where synchronous refers to synchronising the start and end time of hypothesis tests}). In reality, experimentation is ``doubly sequential'' like in Figure~\ref{fig:online_testing}, where it is common to have hypothesis tests that overlap in time, where each test may itself be run sequentially (the \textit{asynchronous} setting).  One natural adjustment for asynchronous testing is to use an online FDR algorithm whenever each test finishes (that is, whichever test is the $t$-th one to finish, test it at level $\alpha_t$). However, this would only assign $\alpha_t$ at the end of a hypothesis test, which would not be appropriate for sequential hypothesis testing and multi-arm bandit approaches that typically require specification of the target type~I error level in advance because it is an important component of their stopping rule. Hence, the testing levels must be specified at the start of a hypothesis test. The asynchronous setting also means that potentially arbitrary dependence between some $p$-values must be considered. Indeed, hypothesis tests that are being conducted concurrently are often likely to be dependent, since they may use the same or highly correlated data during their overlap. 

To address these challenges of asynchronous testing, \citet{Zrnic2018} derived asynchronous versions of LORD++ and SAFFRON that output test levels $\alpha_t$ dynamically at the beginning of the $t$-th test, such that, despite arbitrary local dependence and regardless of the decision times for each hypothesis, the mFDR is controlled at level~$\alpha$. These procedures achieve this goal both at all fixed times $t$, as well as certain adaptively chosen stopping times. \citet{Tian2019} also showed how to derive asynchronous versions of ADDIS. Note that in order to account for the uncertainty about the tests in progress, the test levels assigned by asynchronous online procedures {will often be} more conservative. Thus, there is a trade-off in that although asynchronous procedures take less time to perform a given number of tests, they can be less powerful than their synchronous counterparts.

\citet{Zrnic2019} considered the related setting of online \textit{batched} testing, where a potentially infinite number of batches of hypotheses are tested over time (see Section~\ref{subsec:STAMPEDE} for an example). To this end, they introduced online, FDR-preserving versions of the most widely used offline algorithms, namely the BH procedure and Storey's improvement of the BH method~\citep{storey2002direct}. These online ``mini-batched'' testing algorithms interpolate between online and offline methodology, thus trading off the best of both worlds. {When there is only one batch, the algorithms recover the BH (or Storey-BH) procedure. On the other hand, when all batches are of size one, the algorithms recover the LORD++ (or SAFFRON) procedure}. These algorithms control the FDR under independence (an algorithm valid under positive dependence was also derived), and have a higher power than the fully online testing algorithms. Further, since they consist of compositions of offline FDR algorithms, they imply FDR control over each constituent batch, and not just over the whole sequence of tests. 

\subsubsection{A Bayesian approach}
\citet{gang2021structure} developed a new class of structure-adaptive sequential testing (SAST) rules for online FDR control, which instead of being based on $p$-values, are based upon estimates of the conditional local FDR (Clfdr; \citet{cai2009simultaneous}), which can optimally adapt to important local structures in the data stream. This results in a novel alpha-investing framework that more precisely characterises the effects of rejection of hypotheses: rather than viewing each rejection as a gain of alpha-wealth, the Clfdr characterisation does not view all rejections as equal. Rejections with small Clfdr will lead to increased alpha-wealth whereas rejections with large Clfdr will lead to decreased alpha-wealth. SAST learns the optimal rejection thresholds adaptively and optimises the alpha-wealth allocation across different time periods.
\citet{gang2021structure} showed that SAST can achieve substantial power gain over existing methods, but it comes at the cost of only asymptotically controlling the FDR and requiring the underlying Bayesian model to be well-specified.

\subsubsection{Online FWER control} 
\citet{Tian2021} focused on developing methods for online control of the FWER (see Section~\ref{subsec:error_rates}). Starting with the observation that only alpha-spending had previously been proposed for online FWER control, the authors first extended existing offline algorithms for FWER control, namely the Sidak method~\citep{vsidak1967rectangular} and the fallback procedure~\citep{burman2009recycling}. Given~$T$ hypotheses and FWER level~$\alpha$, the offline Sidak method uses the testing level $1 - (1-\alpha)^{1/T}$ for each hypothesis. The online Sidak method analogously tests hypothesis~$H_t$ at level $\alpha_t = 1 - (1 - \alpha)^{\gamma_t}$. Meanwhile, the offline fallback procedure partitions the overall $\alpha$ between the hypotheses and allows the significance levels to be `recycled' from rejected hypotheses.

The online Sidak and online fallback procedures control the FWER at all times~$t$ under independence and arbitrary dependence, respectively. However, although these online FWER control methods are guaranteed to be uniformly more powerful than alpha-spending, the improvements are usually minor in practice, except for extreme cases. Hence, \citet{Tian2021} proposed the ADDIS-spending algorithm for online FWER control, which (like ADDIS does for online FDR control) benefits from adaptivity to the fraction of nulls, but also gains power by discarding conservative nulls (if they exist). ADDIS-spending controls the FWER when null $p$-values are uniformly conservative, and independent of each other and of the non-nulls. Finally, \citet{fischer2023adaptive} showed how to extend the ADDIS-spending procedure to the setting of `graphical' testing procedures while maintaining control of the FWER. In graphical testing procedures, vertices represent the null hypotheses and weights represent the local significance levels, which are `recycled' through weighted, directed edges. They also showed how to improve the power of the ADDIS-spending procedure under local dependence.

\subsubsection{Online FDX control and FDR at stopping times}
\citet{xu2022dynamic} focused on online control of the FDX (see Section~\ref{subsec:error_rates}). Prior to this work, the only online procedure that controlled the FDX was proposed by~\citet{Javanmard2018}, but this had very low power (no better than alpha-spending). \citet{xu2022dynamic} proposed the supLORD algorithm, which has a higher power and provably controls the FDX when the null $p$-values are conditionally superuniform. One feature of this algorithm is that it allows the user to choose the number of rejections after which FDX control begins, in exchange for more power. supLORD is based on the GAI framework, and the authors also show how to dynamically choose larger test levels $\alpha_t$ when the wealth is large, allowing the algorithm to fully utilise its wealth and increase its power as a result. Finally, the authors show that supLORD also controls the mFDR and FDR at both fixed times and stopping times. Hence, supLORD provides the first guarantee for online FDR control at stopping times (LORD++, SAFFRON and ADDIS only control the mFDR at stopping times and not the FDR).

\subsubsection{Retesting of hypotheses}
One feature of online testing algorithms that has not been explicitly pointed out in the literature is the option of \textit{retesting} hypotheses (i.e.\ using the same $p$-value again later in the testing sequence, when the alpha-wealth may be higher). Crucially however, the choice of whether to retest must be made without using knowledge of the $p$-value itself, but only that it was not rejected (e.g.\ that it is greater 0.01). In this example, under the null the $p$-value will still be conditionally uniform in $[0.01, 1]$. Since this implies that the assumption of conditional super-uniformity under the null still holds, the same $p$-value can be used for retesting. In practice, retesting could happen within an automated testing setting for example, perhaps with additional prior information. Meanwhile, \citet{fisher2022online} proposed a framework for online testing where each hypothesis requires an immediate \textit{preliminary} decision, which allows the analyst to update that decision until a preset deadline while controlling the FDR.

\subsubsection{Discrete test statistics}
\citet{dohler2021online} focused on the setting where the null $p$-values are conservative due to the discreteness of the test statistics, i.e.\ where the individual tests are based on counts or contingency tables. The authors proposed uniform improvements of LORD++, SAFFRON and ADDIS-spending, and showed that the power gains can be substantial when the discreteness is high (e.g.\ the counts in the contingency are moderate).

\subsubsection{Incorporating experimental costs}
{\citet{cook2022cost} considered the setting of online multiple hypothesis testing where the cost of data collection (i.e.\ the cost of conducting an experiment) is not negligible. They proposed an extension of the GAI framework to take into account the cost of data collection, the choice of sample size for each experiment, as well as prior beliefs about the probability of rejection. The proposed methods ensure control of the mFDR and performs particularly well in settings where the aim is to maximise a limited budget of tests to achieve the highest possible power.} 

\subsubsection{Post-hoc FDP bounds}
\citet{katsevich2020simultaneous} proposed a class of simultaneous FDP bounds that apply to a variety of settings, including online testing. These bounds are finite-sample  have a simple closed form. The results can be used as a diagnostic tool for FDR procedures: after running an FDR procedure, one can obtain a valid $1 - \alpha$ confidence bound on the FDP of the resulting rejection set. Since the guarantees are post hoc, they apply to any sequence of rejections produced by any online algorithm, that may or may not have been designed for FDR or FDP control.\\

\subsubsection{Online control of the False Coverage Rate}
Finally, \citet{weinstein2020online} considered the problem of constructing \textit{confidence intervals} (CIs) that are valid for online hypothesis testing. In particular, they focus on control of the false coverage rate (FCR), which is the expected ratio of number of constructed CIs that fail to cover their respective parameters to the total number of constructed CIs. In the online hypothesis testing framework they considered, at each step the investigator observes independent data that are informative about the parameter of interest~$\theta_t$, and must immediately make a decision whether to report a CI for~$\theta_t$ or not. If a CI is reported for $\theta_t$, then the the aim is to ensure that that the CI for $\theta_t$ has $\text{FCR} \leq \alpha$ at all times~$T$. For further details of the proposed algorithms and their theoretical guarantees, see~\citet{weinstein2020online}.

\subsection{Current shortcomings and future directions}
\label{subsec:future}

\noindent \textit{Online testing for small numbers of hypotheses} \\
Online testing algorithms are most powerful in settings where there are a large number (i.e.\ $T > 1000$) of hypotheses that will eventually be tested. Thus the biggest advantage will likely be in settings such as A/B testing in large tech companies or in large-scale biological data repositories. However, while platform trials provide a framework for a trial to continue indefinitely in theory, in practice they will typically evaluate a maximum number of interventions in the low tens.
Hence, there is a need for investigation of optimal online testing procedures when the maximum number of hypotheses to be tested is relatively low and when the correlation between hypotheses is known (e.g.\ because of a shared control arm). Separately,
{there is scope to further improve the power of online testing algorithms when combined with sequential testing of the individual hypotheses, for example by exploiting the fact that pre-specified group-sequential stopping boundaries may be used in a platform trial setting}
(see \citet{zehetmayer2021online} for a recent proposal along these lines).\\

\noindent \textit{Managing incentives across sponsors or products} \\
There are some additional challenges in using online control methods in platform trials or in the IT industry. If different sponsors (e.g.\ pharmaceutical companies) are supporting a platform trial, they might be reluctant to have their intervention be tested at a notably more stringent level than other sponsors: it may be difficult to reconcile the most powerful overall procedure not being acceptable to individual sponsors. Similarly, if a large IT company imposes that experiments run across various products must all be subjected to oversight in the form of a common online FDR controlling procedure that acts across products, then it may be hard to convince individual product teams that their tests must be subject to a level determined by the results of experiments by other groups.\\

\noindent \textit{Optimal choices of parameters for online algorithms} \\
As seen in Section~\ref{sec:methods}, LORD++, SAFFRON and ADDIS depend on the choice of the initial wealth~$w_0$ as well as the sequence $\{\gamma_t\}$. Further work could look at optimal choices of these parameters, given assumptions about the distribution of non-null $p$-values. Exploring data-adaptive choices of time varying sequences $\{\lambda_t\}_{t=1}^{\infty}$ (for SAFFRON and ADDIS) and $\{\eta_t\}_{t=1}^{\infty}$ (for ADDIS) with provable power increase would be another fruitful area of research. Future work could also look at optimal choices for the parameters for the other algorithms in Section~\ref{subsec:further_extensions}. \\

\noindent \textit{Online batched testing} \\
A number of open questions remain regarding the proposals of \citet{Zrnic2019} for online batched testing. First, the framework could be extended to allow for asynchronous online batch testing, using the ideas of~\citet{Zrnic2018}. Second, it should be possible to derive online batched versions based on the offline counterpart of ADDIS, which would gain power in the presence of conservative nulls. Third, an open question is determining the trade off between the chosen batch size versus power in online batched testing.\\ 

\noindent \textit{Online error rate control under dependence} \\
One major shortcoming with almost all of the proposed online testing algorithms is their reliance on the assumption of independence of the null $p$-values for provable FDR control, which is unlikely to always be case in real data applications. However, in terms of online FDR control under dependence, there have only been limited proposals in the literature. \citet{Zrnic2018} showed that the LOND algorithm by~\cite{Javanmard2015} controls the FDR under positive dependence. In the online setting, arbitrary dependence of $p$-values across all time is a rather pessimistic and unrealistic assumption, and thus in the asynchronous setting, \cite{Zrnic2018} introduced the concept of arbitrary \emph{local} dependence and showed that online algorithms can be modified to control the FDR even with such dependence. See also \citet{fisher2021saffron}, who showed further results for control of the FDR under positive dependence in the minibatch setting. Finally, \citet{Zrnic2019} showed how to control the FDR in the online batched setting under positive dependence. Future work could explore how to construct more powerful online algorithms under different forms of dependence, including when the correlation structure is known or estimated.

{\section{Summary and practical guidance \label{sec:summary}}}

\begin{sidewaystable}{
\footnotesize
    \begin{tabular}{l p{4cm} | p{6cm} p{12cm}}
         \textbf{Error rate} & \textbf{Algorithm} & \textbf{Dependence assumptions} & \textbf{Pros \& Cons} \\ \hline
         \textbf{FDR} or \textbf{mFDR} & LORD++ \newline [An online analogue of the BH procedure] & Independence of null $p$-values for FDR control, conditional super-uniformity of null $p$-values for mFDR control & \vspace{-12pt} \begin{itemize}[leftmargin=*]
             \item[+] Extensions for prior weights, penalty weights, decaying memory, as well as local dependence (asynchronous and batch testing) 
             \item[+] Empirically robust to positive dependence of $p$-values
             \item[--] Not robust to arbitrary dependence of $p$-values
             \item[--] Typically lower power than SAFFRON or ADDIS
         \end{itemize} \\[3pt]
         & SAFFRON \newline [Adaptive algorithm based on an estimate of the proportion of true null hypotheses] & As above & \vspace{-12pt}  \begin{itemize}[leftmargin=*]
             \item[+]  Higher power than LORD++ if there is a significant fraction of non-nulls and the signals are strong 
             \item[+] Extensions for local dependence (asynchronous and batch testing)
             \item[--] Not robust to dependence of $p$-values
         \end{itemize}\\[3pt]
         & ADDIS \newline [Combines adaptivity with discarding of conservative nulls] & As above & \vspace{-12pt}  \begin{itemize}[leftmargin=*]
             \item[+] Higher power than SAFFRON when there are conservative nulls
             \item[+] Extensions for local dependence (asynchronous testing)
             \item[--] Not robust to dependence of $p$-values
         \end{itemize} \\[3pt]
         & LOND & Controls FDR under positive dependence of $p$-values & 
         \vspace{-12pt}  \begin{itemize}[leftmargin=*]
             \item[+] Provable FDR control for positive dependence (the `PRDS' assumption)
             \item[--] Substantially lower power than the algorithms above
         \end{itemize} \\[6pt] 
         \textbf{FDX} & supLORD & Null $p$-values are conditionally super-uniform & \vspace{-12pt}  \begin{itemize}[leftmargin=*]
             \item[+] Also controls the mFDR and FDR at both fixed times and stopping times 
             \item[+] User may choose the number of rejections after which we begin controlling FDX in exchange for more power
             \item[--] Unclear how robust to departures from conditional superuniformity
         \end{itemize} \\[6pt]
         \textbf{FWER} & Alpha-spending & --- &  \vspace{-12pt}  \begin{itemize}[leftmargin=*]
             \item[+] Robust to arbitrary dependence of $p$-values 
             \item[--] Very low power, rejects only a few hypotheses before becoming unable to reject any more hypotheses
         \end{itemize} \\[3pt]
         & ADDIS-spending \newline [Combines adaptivity with discarding of conservative nulls] & Null $p$-values are uniformly conservative and independent & \vspace{-12pt}  \begin{itemize}[leftmargin=*]
             \item[+] Higher power than Alpha-spending 
             \item[+] Extensions for local dependence
             \item[--]  Unclear how robust to departures from independence
         \end{itemize} 
    \end{tabular}
    \vspace{6pt}
    \caption{Summary of leading methods for online error rate control, giving dependence assumptions and pros \& cons.}
    \label{tab:summary}
}
\end{sidewaystable}

{Table~\ref{tab:summary} gives a summary of the leading online testing methods discussed in this paper, comparing their assumptions as well as general pros and cons.}
{In terms of practical guidance, we offer the following general suggestions:}\\

\begin{itemize}

\item {A fundamental consideration is which type~I error rate is most suitable to control given the experimental context and goals. As alluded to in Section~\ref{subsec:error_rates}, this choice may be driven by the anticipated number of hypotheses to be tested, data dependencies and/or regulatory concerns.}

\item {Given the type~I error rate that the user wishes to control, there may be a variety of online testing algorithms to choose from. A key consideration is the assumptions around the p-value dependencies, as shown in Table~\ref{tab:STAMPEDE_rej}. Algorithms that make stronger assumptions (i.e., assuming independence) will be more powerful, but this can come at the cost of inflated type~I error rates if these assumptions do not hold. In practice, it may be difficult to anticipate or estimate the data dependencies in an experiment. In some settings, such as a platform trial with a common control, the correlation structure can be derived analytically. Otherwise, with enough data one can try to estimate the correlation empirically.}

\item {The planned timing of hypothesis tests combined with the use of sequential testing may motivate the use of asynchronous or batched versions of online testing algorithms, as discussed in Section~\ref{subsec:further_extensions}. This links with the issue of the ordering of the hypothesis tests themselves: in some settings the ordering will be out of the analyst's control, while in others it may be possible to use prior information about the probability of rejection to potentially gain power either implicitly by ordering the hypotheses (so that those that are a-priori more likely to be rejected are tested first) or by using prior weights (see Section~\ref{subsec:further_extensions}). In the batch setting (i.e., where the multiple hypotheses are available to be tested simultaneously) then the batched algorithms presented in Section~\ref{sec:extensions} are recommended to achieve the best of both worlds of offline and online testing.}

\item {As mentioned above, the setting of a small number of hypotheses ($<1000$) is a challenging one for online testing. Thus the biggest advantage in terms of power will be seen in settings with large-scale hypothesis testing, such as A/B testing. If at some point it becomes known that the number of hypothesis tests will be bounded by a finite number~$M$ then it would make sense to maximise power by ensuring that the alpha-wealth is completely used up by the end of the $M$-th hypothesis test.}

\item {In general, simulation studies remain valuable for assessing the performance of an online testing algorithm given the experimental context and goals, particularly for evaluating power, as well as type~I error rate considerations under departures from independence. We note that in terms of computational scalability, the algorithms presented in this paper all scale linearly with the number of hypotheses tested.}

\end{itemize}


\section{Discussion}
\label{sec:discuss}

Online error rate control methodology provides a powerful and flexible framework for large-scale hypothesis testing that takes into account the temporal nature of modern data analysis. Over the past 15~years since this framework was first proposed, there have been many proposed improvements and extensions, which better reflect the nature of real-world data and expand the scope of potential applications. In particular, continuous progress has been made towards increasing the statistical power of online testing algorithms, so that they can match (and in some cases even exceed) the power of traditional offline algorithms. The issue of accounting for dependent $p$-values remains open, although progress has been made here too. 

As the methodology becomes increasingly mature, the next natural step is to see application of online testing algorithms in practice. To this end, and as seen in Section~\ref{sec:case_studies}, there have been a number of papers that are specifically focused on application examples, including in the context of growing data repositories \citep{robertson2019onlineFDR}, {anomaly detection in time series \citep{rebjock2021online}}, platform trials \citep{robertson2022online} and RNAseq data \citep{liou2022online}. Further work may be required to explore and solve practical challenges that may arise in different application settings. Finally, the provision of software and training will also be key to promoting the use of online error rate control in practice. The \texttt{onlineFDR} package we described earlier is a key step in that regard, but software tuned to specific applications may also be desirable. 

\section*{Acknowledgements}
DS Robertson was supported by the Biometrika Trust, the UK Medical Research Council (MC\_UU\_0002/14) and the NIHR Cambridge Biomedical Research Centre (BRC1215-20014). The views expressed in this publication are those of the authors and not necessarily those of the NHS, the National Institute for Health Research or the Department of Health and Social Care (DHCS). For the purpose of open access, the author has applied a Creative Commons Attribution (CC BY) licence to any Author Accepted Manuscript version arising.
JMS Wason was supported by a NIHR Research Professorship (NIHR301614).
A Ramdas was supported by NSF DMS CAREER award 1916320. \\

\noindent \textit{Data availability statement} \\
The data used in Section~\ref{subsec:IMPC} can be found at \url{https://doi.org/10.5281/zenodo.1343578}. Code to reproduce the simulation studies given in Section~\ref{sec:simulation} can be found at \url{https://github.com/dsrobertson/online_testing}.

%
\begin{appendix}

\section{Test levels for SAFFRON \label{Asec:SAFFRON}}

After choosing $w_0 < \alpha$, {the test levels for} SAFFRON with $\lambda_t \equiv \lambda$ being constant {are defined} as follows:
\begin{enumerate}
\item At each time~$t$, define the number of candidates after the $j$-th rejection as $C_{j+} = C_{j+}(t) = \sum_{i = \tau_j + 1}^{t-1} C_i$, where $C_t= \mathbbm{1}\{ P_t \leq \lambda\}$.

\item SAFFRON starts with $\alpha_1 = \min\{ (1 - \lambda)\gamma_1 w_0, \lambda \}$. Subsequent levels are chosen as $\alpha_t = \min \{ \lambda, \tilde{\alpha}_t \}$, where
\begin{equation*}
\begin{split}
\tilde{\alpha}_t = (1 - \lambda) [w_0 \gamma_{t - C_{0+}} + ( \alpha - w_0) \gamma_{t - \tau_1 - C_{1+}} + \alpha \sum_{j \geq 2}  \, \gamma_{t - \tau_j - C_{j+}} ].
\end{split}
\label{eq:SAFFRON}
\end{equation*}
\end{enumerate}
Formulae for non-constant $\lambda_t$ are in~\citet{Ramdas2018}.

\section{Test levels for ADDIS \label{Asec:ADDIS}}

The testing levels for ADDIS are given by $\alpha_t = \min \{ \lambda, \hat{\alpha}_t\}$, where
\begin{equation*}
\begin{split}
\hat{\alpha}_t = (\eta - \lambda) [w_0 \gamma_{S^t - C_{0+}} + ( \alpha - w_0) \gamma_{S^t - \tau_1^* - C_{1+}} + \alpha \sum_{j \geq 2}  \, \gamma_{S^t - \tau_j^* - C_{j+}} ]
\end{split}
\label{eq:ADDIS}
\end{equation*}
and $S^t = \sum_{i<t} \mathbbm{1} \{ P_i \leq \eta \}$, $\tau_j^* = \sum_{i \leq \tau_j} \mathbbm{1} \{ P_i \leq \eta \}$. 
 See~\citet{Tian2019} for an alternative formulation of ADDIS where $p$-values greater than $\eta$ are explicitly discarded, {and the extension} to a sequence $\{ \eta_t\}_{t=1}^{\infty}$.

{\section{Simulation study \label{Asec:sim}}}

\begin{figure}[ht!]
\centering
\includegraphics[width=0.85\linewidth]{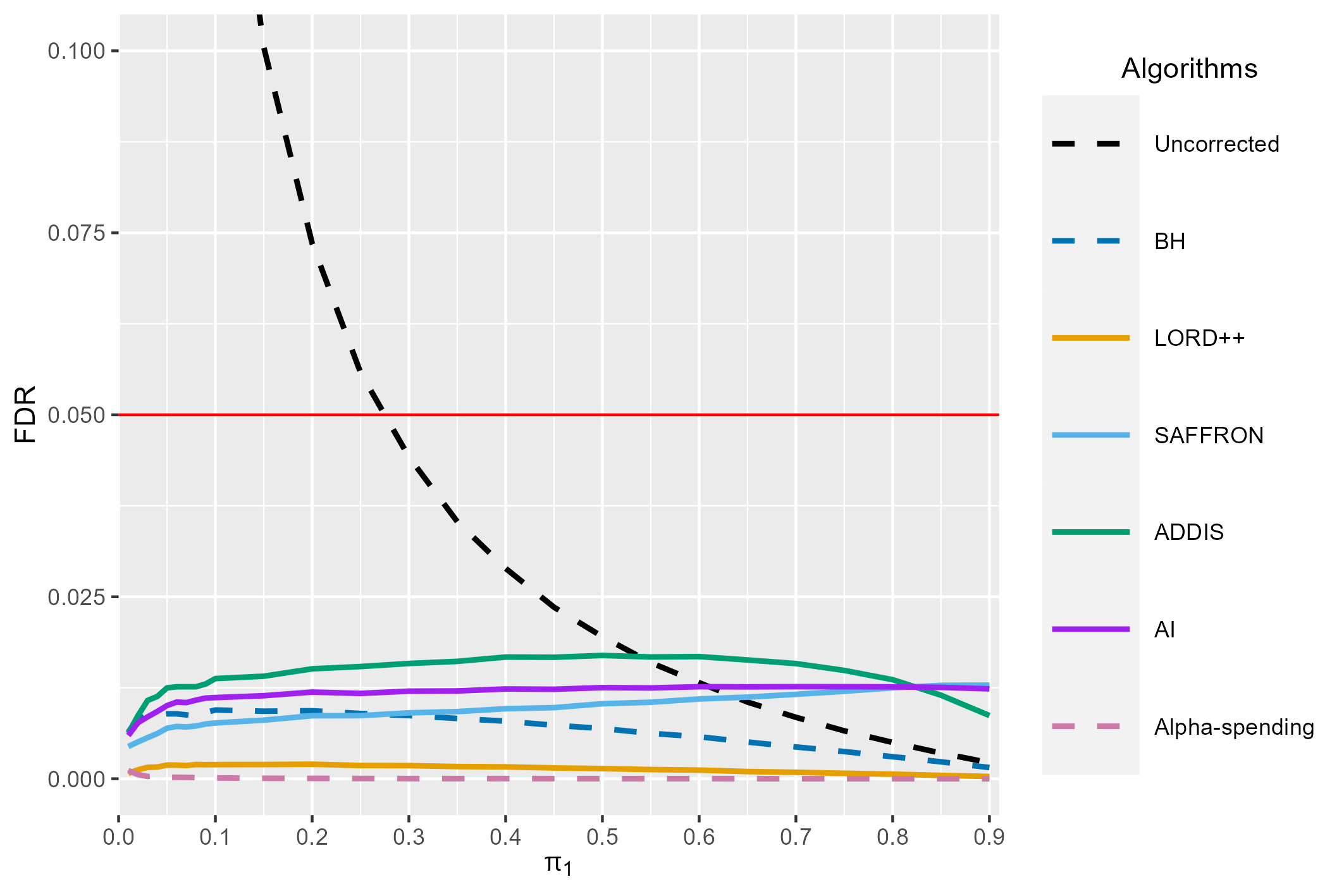}
\caption{FDR of LORD++, SAFFRON, ADDIS {and monotone AI} compared with uncorrected testing, the BH procedure and alpha-spending as the proportion of non-nulls $\pi_1$ varies. The solid red horizontal line gives the target level of $\alpha = 0.05$. We set $T = 1000$ and results are based on $10^4$ simulation replicates.}
\label{fig:FDR}
\end{figure}

Figure~\ref{fig:FDR} shows the FDR of LORD++, SAFFRON, ADDIS and monotone AI compared with uncorrected testing, the BH procedure and alpha-spending, using the simulation set-up described in Section~\ref{subsec:test_gaussian}.

\end{appendix}


\bibliography{bibliography}       


\end{document}